\documentstyle[12pt]{article}

\begin{document}
\input{psfig.sty}
\begin{flushright}
\baselineskip=12pt
UPR-953-T \\
\end{flushright}

\begin{center}
\vglue 1.5cm
{\Large\bf N=2 Supersymmetric GUT Breaking on $T^2$ Orbifolds}
\vglue 2.0cm
{\Large Tianjun Li~\footnote{E-mail: tli@bokchoy.hep.upenn.edu,
phone: (215) 898-7938, fax: (215) 898-2010.}}
\vglue 1cm
{ Department of Physics and Astronomy \\
University of Pennsylvania, Philadelphia, PA 19104-6396 \\  
U.  S.  A.}
\end{center}

\vglue 1.5cm
\begin{abstract}
We study the 6-dimensional 
$N=2$ supersymmetric grand unified theories with gauge
group $SU(N)$ and $ SO(M)$ on the extra space orbifolds $T^2/(Z_2)^3$ and
$T^2/(Z_2)^4$, which can be broken down to the 4-dimensional $N=1$
supersymmetric
$SU(3)\times SU(2)\times U(1)^{n-3}$
model for the zero modes. We also 
study the models which have two $SU(2)$ Higgs doublets 
(zero modes) from the 6-dimensional vector multiplet. 
We give the particle spectra, present the fields, the number of
4-dimensional
supersymmetry and gauge group on the observable 3-brane or
4-brane and discuss some phenomenology for those models. Furthermore,
we generalize our procedure for $(4+m)$-dimensional $N$ supersymmetric
GUT breaking on the space-time
 $M^4\times T^m/(Z_2)^L$.
\\[1ex]
\\[1ex]
\\[1ex]
Los Alamos Database Number: hep-ph/0108120

\end{abstract}

\vspace{0.5cm}
\begin{flushleft}
\baselineskip=12pt
August 2001\\
\end{flushleft}
\newpage
\setcounter{page}{1}
\pagestyle{plain}
\baselineskip=14pt

\section{Introduction}
Grand unified theory (GUT) gives us an simple
 and elegant understanding of the quantum numbers of the quarks and
leptons,
and the success of gauge coupling unification in the Minimal
Supersymmetric
Standard Model strongly support
 this idea. Although the GUT at high energy scale has
become widely accepted now, there are some problems in GUT:  the grand
unified
gauge symmetry breaking mechanism, the doublet-triplet splitting problem
and
 proton decay.

A new scenario to explain above questions in GUT has been suggested 
by Kawamura~\cite{kaw1, kaw2, kaw3},
 and further discussed by Altarelli, Barbieri, Feruglio, Hall,
Hebecker, Kawamoto,
Normura, and March-Russell~\cite{AF, HN, kk, HMR, bhn1N, bhn1, HMRN}.
 The key point is that the GUT
gauge symmetry exists in 5 or higher dimensions and is broken down to the
Standard Model gauge symmetry for the zero modes by non-trivial 
orbifold projection on the multiplets and gauge generators in GUT. 
The attractive models have been constructed explicitly, where
the supersymmetric 5-dimensional SU(5) models are broken down to
the N=1 supersymmetric Standard Model through the compactification on
$S^1/(Z_2\times Z_2')$. The GUT symmetry breaking and doublet-triplet
splitting problems have been solved neatly by the orbifold projection.
Recently, gauge symmetry breaking and supersymmetry breaking
due to the orbifold projection
or the Scherk-Schwartz mechanism have been discussed by a lot of 
papers~[11-21]. By the way, it has been discussed previously that
the gauge symmetry and supersymmetry can be broken due to
the Scherk-Schwartz mechanism in string theory with large extra 
dimensions~\cite{IA}.

In this paper, we would like to study the $N=2$ supersymmetric
GUT breaking on the space-time $M^4\times T^2$ where
$M^4$ is the ordinary 4-dimensional Minkowski space-time.
 We assume that the grand unified 
theory is the higher dimensional theory and
valid above the $T^2$ compactification scale, i. e.,
at very high energy scale or temperature. When the energy
scale or temperature goes down, the extra space orbifold becomes
compact. And then, the fields in GUT, which do not have the zero modes
due to the orbifold projections, obtain the masses proportional
to the $T^2$ compactification scale and decouple to the 4-dimensional
low energy theory. Only the fields, which have the zero modes, remain in
the 4-dimensional low energy theory. So, the 
4-dimensional low energy theory may not
preserve the original GUT gauge symmetry and supersymmetry.
We show that the 6-dimensional
 $N=2$ supersymmetric GUT models can be broken down to
the 4-dimensional $N=1$ supersymmetric $SU(3)\times SU(2) \times
U(1)^{n-3}$
models for the zero modes in the bulk, where $n$ is the rank of GUT
group.
In addition, if the Higgs triplets did not have the zero modes and
the Higgs doublets did have the zero modes, we can solve the 
doublet-triplet splitting problem for the masses of Higgs
triplets are proportional to the $T^2$ compactification scale.
 And if the $T^2$ compactfication scale is
about $10^{16}$ GeV, the proton decay problem can also be solved.
In short, we can break the GUT gauge symmetry and supersymmetry at
low energy scale, and
solve the doublet-triplet splitting problem and proton decay problem
in this kind of approach.

As we know, $N=1$ 6-dimensional supersymmetric theory is
chiral, where the gaugino (and gravitino) has positive chirality
and the matters (hypermultiplets) have negative chirality, so, 
it often has anomaly. In order to avoid anomaly, we can consider
the models with no supersymmetry~\cite{TJ} or the models with $N=2$
supersymmetry 
in the bulk~\cite{HNDS, ABC, HNODS}. 
For the first kind of scenario, we put the gauge field and/or
Higgs fields in the bulk, and the SM fermions on the observable brane.
 In this kind
of scenario, on the observable brane, the SU(5) gauge symmetry is broken
down
to the $SU(3) \times SU(2) \times U(1) $ gauge symmetry due to
the non-trivial orbifold projection, and
 there may exist only the zero modes and KK modes
of the Standard Model gauge fields and Higgs fields
because we project out not only the zero modes of the 
non-Standard Model gauge fields and triplet Higgs fields, 
but also their KK modes. So, the extra dimensions
can be large and the gauge hierarchy problem can be solved 
for there does not exist proton decay problem at all.
The second kind of scenario is the subject of 
this paper. Our convention is given in section 2.
 Because $N=2$ 6-dimensional supersymmetric theory 
has 16 real supercharges,
which corresponds to $N=4$ 4-dimensional supersymmetric theory,
there are no hypermultiplets in the bulk. We assume that
only the gauge fields are in the bulk, the SM model fermions
and Higgs particles are on the observable 3-brane
which is located at the fixed 
point, or on the observable 4-brane which is located 
at the fixed line (boundary) in the extra space orbifold.
In terms of 4-dimensional language, $N=4$ vector
multiplets have a vector multiplet $V_{\mu}$, and
three chiral multiplets $\Sigma_5$, $\Sigma_6$ and $\Phi$.

First, we will discuss the  models without gauge-Higgs unification.
We require that: (1) there are no zero modes for the
chiral multiplets $\Sigma_5$, $\Sigma_6$ and $\Phi$;
(2) the zero modes for $V_{\mu}$ preserve $N=1$ supersymmetry and
$SU(3)\times SU(2)\times U(1)^{n-3}$ gauge symmetry where $n$ is the
rank of the GUT gauge group. So,
 we introduce two $Z_2$ parities $P^y$ and
$P^z$, which are unit matrix in the adjoint representation
of the GUT gauge group. Considering the zero modes,
under $P^y$ projection, we can break the 4-dimensional $N=4$ supersymmetry
to
$N=2$ supersymmetry with $(V, \Sigma_6)$ forming a vector multiplet and
$(\Sigma_5, \Phi)$ forming a hypermultiplet, and we can break the
4-dimensional $N=2$ supersymmetry to $N=1$ supersymmetry further by
$P^z$ projection. Moreover, we use additional parity operators $P^{y'}$,
$P^{z'}$ and
$P^{y' z'}$ to break the GUT gauge symmetry.
Explicitly, we will discuss two N=2
supersymmetric $SU(5)$ models on the space-time
$M^4\times T^2/(Z_2)^3$: one is on
 $M^4\times (S^1/Z_2 \times S^1/Z_2)/Z_2$, the other is
on $M^4\times S^1/(Z_2)^2 \times S^1/Z_2$. 
We also discuss the
supersymmetric $SU(5)$, $SU(6)$ and $SO(10)$ models on the
space-time $M^4\times T^2/(Z_2)^4$ by orbifold projections.
For those models, we calculate the mass spectra, 
list the superfields, the number of the corresponding 4-dimensional
supersymmetry, and the gauge group on the observable 3-brane or
4-brane.
The fixed lines are fixed under one projection, so, including 
the KK modes, 
the 4-branes preserve 4-dimensional $N=2$ supersymmetry.
And the fixed points are usually fixed under two projections, or in
other words, two 4-branes' intersection, so, the 3-branes preserve 
 4-dimensional $N=1$ supersymmetry, except for the
orbifold $T^2/(Z_2)^3 = (S^1/Z_2\times S^1/Z_2)/Z_2$, where
the vertex is the fixed point under the projection $(y', z') \sim (-y',
-z')$,
so, the 3-brane at that point preserves 4-dimensional $N=2$
supersymmetry.

Moreover, we discuss the gauge-Higgs unification.
In this kind of model, we require that:
 (1) there are no zero modes for the chiral multiplets $\Sigma_5$ 
and $\Sigma_6$, and for the zero modes, the 4-dimensional
$N=4$ GUT symmetry is broken down to the $N=1$ $SU(3)\times SU(2)\times
U(1)^{n-3}$
gauge symmetry; 
(2) considering the zero modes,
there exist only one pair of Higgs doublets from chiral multiplet
$\Phi$, because if
we had two pairs of Higgs doublets, we may have the flavour changing
neutral current problem. We discuss the 
$SU(6)$ models on the space-time $M^4\times T^2/(Z_2)^3$ completely.
We also discuss the $SU(6)$, $SU(7)$ and $SO(12)$ on the space-time 
$M^4\times T^2/(Z_2)^4$. For all of those models, 
we calculate the superfields, the number of the corresponding
4-dimensional
supersymmetry, and the gauge group on the 3-brane and
4-brane. We also comment on the $SU(7)$ and $SO(12)$ models on the
space-time
$M^4\times T^2/(Z_2)^3$ with zero modes from chiral multiplets $\Sigma_5$
and $\Sigma_6$.

Furthermore, we discuss some phenomenology
of those models. We point out that
for rank $n$ semi-simple group, after those $Z_2$ projections,
the minimal gauge group will be $SU(3)\times SU(2) \times U(1)^{n-3}$
because we can not project out the gauge generators in the Cartan
subalgebra.
So, we have to break the additional $U(1)$ symmetry by introducing
extra chiral superfields. CP violations can be introduced if
we considered the complex $F$-term SUSY breaking~\cite{HLLY}.
 And proton decay can be avoided 
by introducing suitable $R$ charges to the fields due to $R$ symmetry, 
and by giving the triplet Higgs masses about 
 $1/R_1$ or $1/ R_2$ at the order of GUT scale through the projections.

We also generalize our procedure for $(4+m)$-dimensional $N$
supersymmetric
GUT with gauge group $G$ breaking on the space-time $M^4\times
T^m/(Z_2)^L$
where $L~\le~ 2m$. Because
we can have more $Z_2$ symmetries by introducing more extra dimensions,
we can discuss the higher rank GUT gauge symmetry breaking on the higher
dimensional space-time.

\section{$N=2$ 6D Supersymmetric Gauge Theory and Convention}
We consider the 6-dimensional gauge theory with $N=2$ supersymmetry.
$N=2$ supersymmetric theory in 6-dimension has 16 real supercharges,
corresponding to $N=4$ supersymmetry in 4-dimension. So, only the
vector multiplet can be introduced in the bulk. 
In terms of 4-dimensional
$N=1$ language, it contains a vector multiplet $V(A_{\mu}, \lambda_1)$,
and three chiral multiplets $\Sigma_5$, $\Sigma_6$, and $\Phi$. All 
of them are in the adjoint representation of the gauge group. In addition,
the $\Sigma_5$ and $\Sigma_6$ chiral multiplets
contain the gauge fields $A_5$ and $A_6$ in
their lowest components, respectively.

In the Wess-Zumino gauge and 4-dimensional $N=1$ language, the bulk action 
is~\cite{NAHGW}
\begin{eqnarray}
  S &=& \int d^6 x \Biggl\{
  {\rm Tr} \Biggl[ \int d^2\theta \left( \frac{1}{4 k g^2} 
  {\cal W}^\alpha {\cal W}_\alpha + \frac{1}{k g^2} 
  \left( \Phi \partial_5 \Sigma_6 - \Phi \partial_6 \Sigma_5
  - \frac{1}{\sqrt{2}} \Phi 
  [\Sigma_5, \Sigma_6] \right) \right) 
\nonumber\\
&& + {\rm H.C.} \Biggr] 
  + \int d^4\theta \frac{1}{k g^2} {\rm Tr} \Biggl[ 
  \sum_{i=5}^6 \left((\sqrt{2} \partial_i + \Sigma_i^\dagger) e^{-V} 
  (-\sqrt{2} \partial_i + \Sigma_i) e^{V} + 
   \partial_i e^{-V} \partial_i e^{V}\right)
\nonumber\\
  && \qquad \qquad \qquad
  + \Phi^\dagger e^{-V} \Phi e^{V}  \Biggr] \Biggr\} ~.~\,
\label{eq:5daction}
\end{eqnarray}
And the gauge transformation is given by
\begin{eqnarray}
  e^V &\rightarrow& e^\Lambda 
    e^V e^{\Lambda^\dagger}, \\
  \Sigma_i &\rightarrow& e^\Lambda (\Sigma_i - \sqrt{2} \partial_i) 
    e^{-\Lambda}, \\
  \Phi &\rightarrow& e^\Lambda \Phi e^{-\Lambda}~,~\,
\end{eqnarray}
where $i=5, 6$.

We would like to explain our convention. We consider 
the 6-dimensional space-time which can be factorized into a product of the 
ordinary 4-dimensional Minkowski space-time $M^4$, and the torus $T^2$
which is homeomorphic to $S^1\times S^1$. The corresponding
coordinates for the space-time are $x^{\mu}$, ($\mu = 0, 1, 2, 3$),
$y\equiv x^5$ and $z\equiv x^6$. 
The radii for the circles along $y$ direction and $z$ direction are
$R_1$ and $R_2$, respectively.
We can also define $y'$ and $z'$ by $y' \equiv y-\pi R_1/2$ 
and $z' \equiv z- \pi R_2/2$. In addition,
 we assume that the gauge fields are in
the bulk and the SM fermions and Higgs particles are on the observable 
4-brane which can be located at the 
 fixed line (boundary) or on the observable 3-brane which
can be located at the fixed point
in the extra space orbifold. By the way, we define the observable brane
as the brane which contains the Standard Model fermions.

The orbifold $T^2/Z_2$ can be obtained by
moduloing three different kinds of equivalent classes:
\begin{equation}
(1)~ y\sim -y~;~(2)~ z\sim -z ~;~(3)~  (y, z) \sim (-y, -z)~,~\,
\end{equation}
which exactly speaking correspond to $S^1/Z_2\times S^1$, $S^1\times
S^1/Z_2$
and $T^2/Z_2$, respectively.

And the orbifold $T^2/(Z_2)^2$ can be obtained by moduloing six
different kinds of equivalent classes
\begin{equation}
 y\sim -y~,~y'\sim -y'~,~\,
\end{equation}
\begin{equation}
 z\sim -z~,~z'\sim -z'~,~\,
\end{equation} 
\begin{equation}
 y\sim -y~,~z\sim -z~,~\,
\end{equation}
\begin{equation}
 (y, z)\sim (-y, -z)~,~(y', z')\sim (-y', -z')~,~\,
\end{equation}
\begin{equation}
 (y, z)\sim (-y, -z)~,~y'\sim -y'~,~\,
\end{equation}
\begin{equation}
 (y, z)\sim (-y, -z)~,~z'\sim -z'~,~\,
\end{equation}
which exactly speaking correspond to $S^1/(Z_2)^2\times S^1$, $S^1\times
S^1/(Z_2)^2$,
$S^1/Z_2\times S^1/Z_2$, $(T^2/Z_2)/Z_2$, $(S^1/Z_2\times S^1)/Z_2$,
$(S^1\times S^1/Z_2)/Z_2$,
respectively. By the way,  because the translations:
$y\longrightarrow y \pm \pi R_1/2$ and $z\longrightarrow z \pm \pi R_2/2$,
are invariant in the covering space 
$S^1\times S^1$ or $R^1\times R^1$ if we pull back,
we consider $\{y\sim -y,~(y',  z)\sim (-y', - z)\}$ 
and $\{z \sim -z,~( y, z')\sim ( -y, -z')\}$ are equivalent to
$\{(y, z)\sim (-y, -z),~y'\sim -y'\}$ and $\{(y, z)\sim (-y, -z)~,~z'\sim
-z'\}$, respectively.

In addition, the orbifold $T^2/(Z_2)^3$ can also be obtained by moduloing
three
different kinds of equivalent classes
\begin{equation}
 y\sim -y~,~ z\sim -z~,~ y'\sim -y'~,~\,
\end{equation}
\begin{equation}
y\sim -y~,~ z\sim -z~,~z'\sim -z'~,~\,
\end{equation} 
\begin{equation}
 y\sim -y~,~ z\sim -z ~,~(y', z')\sim (-y', -z')~,~\,
\end{equation}
which exactly speaking correspond to $S^1/(Z_2)^2\times S^1/Z_2$,
$S^1/Z_2\times S^1/(Z_2)^2$,
and $(S^1/Z_2\times S^1/Z_2)/Z_2$, respectively. By the
way, we consider $\{(y, z)\sim (-y, -z),~ y' \sim -y',~z'\sim -z'\}$
is equivalent to $\{y\sim -y,~ z\sim -z,~(y', z')\sim (-y', -z')\}$
due to the translation 
($y\longrightarrow y \pm \pi R_1/2$ and $z\longrightarrow z \pm \pi
R_2/2$)
 invariant in the covering space.

Moreover, there exists one $T^2/(Z_2)^4$ by moduloing the
equivalent classes
\begin{equation}
 y\sim -y~,~ z\sim -z~,~ y'\sim -y'~,~ z'\sim -z'~.~\,
\end{equation}

By the way, one may ask whether one can define  $y''$ and $z''$ by $y''
\equiv y-\pi R_1/4$ 
and $z'' \equiv z - \pi R_2/4$, and introduce the equivalent classes, for
example, like $y'' \sim -y''$, $z'' \sim -z''$ or $(y'', z'') \sim (-y'',
-z'')$.
However, the extra $Z_2''$ symmetry is
equivalent to the $Z_2$ or $Z_2'$ symmetry because we can have at most
two non-equivalent $Z_2$ reflection symmetries
 along one extra dimension~\cite{TJL}. 

For a generic bulk field $\phi(x^{\mu}, y, z)$,
we can define six parity operators $P^y$, $P^z$, $P^{y z}$,
$P^{y'}$, $P^{z'}$, and $P^{y' z'}$ (only 4 of them are independent),
respectively,
\begin{eqnarray}
\phi(x^{\mu},y, z)&\to \phi(x^{\mu},-y, z )=P^y \phi(x^{\mu},y, z)
 ~,~\,
\end{eqnarray}
\begin{eqnarray}
\phi(x^{\mu},y, z)&\to \phi(x^{\mu}, y, -z )=P^z \phi(x^{\mu},y, z)
 ~,~\,
\end{eqnarray}
\begin{eqnarray}
\phi(x^{\mu},y, z)&\to \phi(x^{\mu},-y, -z )= P^{y z} \phi(x^{\mu},y, z)
 ~,~\,
\end{eqnarray}
\begin{eqnarray}
\phi(x^{\mu},y', z')&\to \phi(x^{\mu},-y', z' )=P^{y'} \phi(x^{\mu},y',
z')
 ~,~\,
\end{eqnarray}
\begin{eqnarray}
\phi(x^{\mu},y', z')&\to \phi(x^{\mu}, y', -z' )=P^{z'} \phi(x^{\mu},y',
z')
 ~,~\,
\end{eqnarray}
\begin{eqnarray}
\phi(x^{\mu},y', z')&\to \phi(x^{\mu},-y', -z' )= P^{y' z'}
\phi(x^{\mu},y', z')
 ~.~\,
\end{eqnarray}

From the bulk action, we obtain that
under the transformations of $P^y$, $P^z$ and $P^{y z}$, the
vector multiplet transforms as
\begin{eqnarray}
  V(x^{\mu},-y, z) &=& P^y V(x^{\mu}, y, z) (P^y)^{-1}
~,~\,
\end{eqnarray}
\begin{eqnarray}
  \Sigma_5(x^{\mu},-y, z) &=& - P^y \Sigma_5(x^{\mu}, y, z) (P^y)^{-1}
~,~\,
\end{eqnarray}
\begin{eqnarray}
  \Sigma_6(x^{\mu},-y, z) &=&  P^y \Sigma_6(x^{\mu}, y, z) (P^y)^{-1}
~,~\,
\end{eqnarray}
\begin{eqnarray}
  \Phi(x^{\mu},-y, z) &=& - P^y \Phi(x^{\mu}, y, z) (P^y)^{-1}
~,~\,
\end{eqnarray}
\begin{eqnarray}
  V(x^{\mu},y, -z) &=& P^z V(x^{\mu}, y, z) (P^z)^{-1}
~,~\,
\end{eqnarray}
\begin{eqnarray}
  \Sigma_5(x^{\mu}, y, -z) &=&  P^z \Sigma_5(x^{\mu}, y, z) (P^z)^{-1}
~,~\,
\end{eqnarray}
\begin{eqnarray}
  \Sigma_6(x^{\mu}, y, -z) &=&  - P^z \Sigma_6(x^{\mu}, y, z) (P^z)^{-1}
~,~\,
\end{eqnarray}
\begin{eqnarray}
  \Phi(x^{\mu}, y, -z) &=& - P^z \Phi(x^{\mu}, y, z) (P^z)^{-1}
~,~\,
\end{eqnarray}
\begin{eqnarray}
  V(x^{\mu}, -y, -z) &=& P^{y z} V(x^{\mu}, y, z) (P^{y z})^{-1}
~,~\,
\end{eqnarray}
\begin{eqnarray}
  \Sigma_5(x^{\mu}, -y, -z) &=& - P^{y z} \Sigma_5(x^{\mu}, y, z)
 (P^{y z})^{-1}
~,~\,
\end{eqnarray}
\begin{eqnarray}
  \Sigma_6(x^{\mu}, -y, -z) &=&  - P^{y z} \Sigma_6(x^{\mu}, y, z) 
(P^{y z})^{-1}
~,~\,
\end{eqnarray}
\begin{eqnarray}
  \Phi(x^{\mu}, -y, -z) &=&  P^{y z} \Phi(x^{\mu}, y, z) (P^{y z})^{-1}
~.~\,
\end{eqnarray}
And under parity operators $P^{y'}$, $P^{z'}$ and $P^{y' z'}$, the
vector multiplet transformations are similar to those under
$P^{y}$, $P^{z}$ and $P^{y z}$.

In addition, for group $G$ with unit element $e$ in the adjoint
representation, 
we define 
\begin{eqnarray}
G/{P^{y'}}~\equiv~ G/\{e, P^{y'}\} ~,~\,
\end{eqnarray}
\begin{eqnarray}
G/\{P^{y'} \cup P^{z'}\}~\equiv~ G/\{e, P^{y'}\} \cap G/\{e, P^{z'}\}
 ~,~\,
\end{eqnarray}
similar for the others.

\section{ N=2 Supersymmetric $SU(5)$ on $M^4 \times T^2/(Z_2)^3$}

In this section, we would like to discuss $N=2$ supersymmetric
$SU(5)$ model on the space-time $M^4 \times T^2/(Z_2)^3$ without
gauge-Higgs unification. We require that: (1) there are no zero modes for
chiral multiplets $\Sigma_5$, $\Sigma_6$ and $\Phi$;
(2) for the zero modes, we only have
4-dimensional $N=1$ $SU(3)\times SU(2)\times U(1)$
gauge symmetry.

We will discuss $SU(5)$ models on various orbifolds in
detail, because it is easy to generalize our procedure to other
higher rank gauge group $SU(N)$ ($N > 5$), $SO(M)$ ($M\ge 10$), 
$E_6$, $E_7$ and $E_8$ by introducing more $Z_2$ projections.
We will choose the matrix representations
in the adjoint representation for $P^y$ and $P^z$ as unit matrix,
i. e., ${\rm diag} (+1, +1, +1, +1, +1)$. Considering the zero modes,
under $P^y$ projection, we can break the 4-dimensional $N=4$ supersymmetry
to
$N=2$ supersymmetry with $(V, \Sigma_6)$ forming a vector multiplet and
$(\Sigma_5, \Phi)$ forming a hypermultiplet, and we can break the
4-dimensional $N=2$ supersymmetry to $N=1$ supersymmetry further by
$P^z$ projection. 

\subsection{$SU(5)$ on $M^4\times (S^1/Z_2\times S^1/Z_2)/Z_2$}

First, we will consider the $T^2/(Z_2)^3$ obtained by 
$T^2$ moduloing the equivalent classes: $y \sim -y$, $ z \sim -z$,
$(y', z') \sim (-y', -z')$. 
 The three non-equivalent fixed points 
at which the observable 3-brane can be located are
$(y=0, z=0)$, $(y=0, z=\pi R_2)$ and $(y=\pi R_1/2, z=\pi R_2/2)$,
where the first two points are fixed
under the first two projections ($y\sim -y$ and $z\sim -z$) and the
last one is fixed under the last projection 
($(y', z')\sim (-y', -z')$). Moreover, the two fixed lines at which
the observable 4-brane can be located are $y=0$ and $z=0$.
The extra space orbifold is
$(S^1/Z_2\times S^1/Z_2)/Z_2$, which is a cone with
vertex $(y=\pi R_1/2, z=\pi R_2/2)$ and two boundaries 
$\{(y, z)| y=0, z \in [0, \pi R_2]\}$ and
$\{(y, z)| y\in [0, \pi R_1], z=0 \}$.

For a generic bulk field $\phi(x^{\mu}, y, z)$,
we can define three parity operators $P^y$, $P^z$ and $P^{y' z'}$,
respectively.
Denoting the field with ($P^y$, $P^z$, $P^{y' z'}$)=($\pm, \pm, \pm$) by 
$\phi_{\pm \pm \pm}$, we can obtain the KK mode expansions, which are
given in 
Appendix A.

We choose the following matrix representations for parity operators 
$P^y$, $P^z$ and $P^{y' z'}$, 
which are expressed in the adjoint representation of SU(5)
\begin{equation}
P^y={\rm diag}(+1, +1, +1, +1, +1)
~,~ P^z={\rm diag}(+1, +1, +1, +1, +1)~,~\,
\end{equation}
\begin{equation}
P^{y' z'}={\rm diag}(-1, -1, -1, +1, +1)
 ~.~\,
\end{equation}

In addition, under $P^{y' z'}$ parity,
the $SU(5)$ gauge generators $T^A$ where A=1, 2, ..., 24 for SU(5)
are separated into two sets: $T^a$ are the gauge generators for
the Standard Model gauge group, and $T^{\hat a}$ are the other broken
gauge generators 
\begin{equation}
P^y~T^a~(P^y)^{-1}= T^a ~,~ P^y~T^{\hat a}~(P^y)^{-1}= T^{\hat a}
~,~\,
\end{equation}
\begin{equation}
P^z~T^a~(P^z)^{-1}= T^a ~,~ P^z~T^{\hat a}~(P^z)^{-1}= T^{\hat a}
~,~\,
\end{equation}
\begin{equation}
P^{y' z'}~T^a~(P^{y' z'})^{-1}= T^a ~,~ P^{y' z'}~T^{\hat a}
~(P^{y' z'})^{-1}= - T^{\hat a}
~.~\,
\end{equation}

\renewcommand{\arraystretch}{1.4}
\begin{table}[t]
\caption{Parity assignment and masses ($n\ge 0, m \ge 0$) for
the vector multiplet in the $SU(5)$ model on 
$(S^1/Z_2\times S^1/Z_2)/Z_2$.
The index $a$ labels the unbroken $SU(3)\times SU(2) \times U(1)$ 
gauge generators, while $\hat a$
labels the other broken SU(5) gauge generators.
\label{tab:SUV1}}
\vspace{0.4cm}
\begin{center}
\begin{tabular}{|c|c|c|}
\hline        
$(P^y, P^z, P^{y' z'})$ & field & mass\\ 
\hline
$(+, +, +)$ &  $V^a_{\mu}$ & $\sqrt {(2n)^2/R_1^2+ (2m)^2/R_2^2}$ or 
$\sqrt {(2n+1)^2/R_1^2+ (2m+1)^2/R_2^2}$\\
\hline
$(+, +, -)$ &  $V^{\hat{a}}_{\mu}$ & $\sqrt {(2n+1)^2/R_1^2+(2m)^2/R_2^2}$
or $\sqrt {(2n)^2/R_1^2+(2m+1)^2/R_2^2}$ \\
\hline
$(-, +, -)$ &  $\Sigma_5^a$ & $\sqrt {(2n+2)^2/R_1^2+ (2m)^2/R_2^2}$ or 
$\sqrt {(2n+1)^2/R_1^2+ (2m+1)^2/R_2^2}$\\
\hline
$(-, +, +)$ &  $\Sigma_5^{\hat{a}}$ & $\sqrt {(2n+1)^2/R_1^2+
(2m)^2/R_2^2}$ or 
$\sqrt {(2n+2)^2/R_1^2+ (2m+1)^2/R_2^2}$\\
\hline
$(+, -, -)$ &  $\Sigma_6^a$ & $\sqrt {(2n)^2/R_1^2+ (2m+2)^2/R_2^2}$ or 
$\sqrt {(2n+1)^2/R_1^2+ (2m+1)^2/R_2^2}$\\
\hline
$(+, -, +)$ &  $\Sigma_6^{\hat{a}}$ & $\sqrt {(2n)^2/R_1^2+
(2m+1)^2/R_2^2}$ or 
$\sqrt {(2n+1)^2/R_1^2+ (2m+2)^2/R_2^2}$\\
\hline
$(-, -, +)$ &  $\Phi^a$ & $\sqrt {(2n+1)^2/R_1^2+ (2m+1)^2/R_2^2}$ or 
$\sqrt {(2n+2)^2/R_1^2+ (2m+2)^2/R_2^2}$\\
\hline
$(-, -, -)$ &  $\Phi^{\hat{a}}$ & $\sqrt {(2n+2)^2/R_1^2+(2m+1)^2/R_2^2}$
or $\sqrt {(2n+1)^2/R_1^2+(2m+2)^2/R_2^2}$ \\
\hline
\end{tabular}
\end{center}
\end{table}

\renewcommand{\arraystretch}{1.4}
\begin{table}[t]
\caption{For the model $SU(5)$ on
$(S^1/Z_2\times S^1/Z_2)/Z_2$,
the gauge superfields, the number of 4-dimensional supersymmetry
 and gauge symmetry on the 3-brane, which
is located at the fixed point $(y=0, z=0),$ $(y=0, z=\pi R_2),$  and 
$(y=\pi R_1/2, z=\pi R_2/2)$, and on the 4-brane which is located at 
the fixed line $y=0$ and $z=0$.
\label{tab:SUV11}}
\vspace{0.4cm}
\begin{center}
\begin{tabular}{|c|c|c|c|}
\hline        
Brane Position & Fields & SUSY & Gauge Symmetry\\ 
\hline
$(0, 0) $ &  $V^A_{\mu}$ & $N=1$ & $SU(5)$ \\
\hline
$(0, \pi R_2)$ & $V^A_{\mu}$ & N=1 & $SU(5)$ \\
\hline
$(\pi R_1/2, \pi R_2/2) $ & $V^a_{\mu}$, $\Sigma_5^{\hat a}$,
$\Sigma_6^{\hat a} $, $\Phi^a$
  & N=2 & $SU(3)\times SU(2)\times U(1)$ \\
\hline
$y=0$ & $V^A_{\mu}$, $\Sigma_6^A$  & N=2 & $SU(5)$ \\
\hline
$z= 0 $ & $V^A_{\mu}$, $\Sigma_5^A$  & N=2 & $SU(5)$ \\
\hline
\end{tabular}
\end{center}
\end{table}

It is not difficult to obtain the particle spectra, which are given in
 Table 1. We also present the gauge superfields, the number of
4-dimensional
supersymmetry, and gauge group on the 3-brane or 4-brane in Table 2.
Usually, the fixed lines are fixed under one projection, and
the fixed points are fixed under two projections if they were at 
two fixed lines' intersections,
 so, including the KK modes, 
 the number of 4-dimensional supersymmetry is $N=1$ on the
 3-branes at the fixed points,
and $N=2$ on the 4-branes
at the fixed lines. However,  $(y=\pi R_1/2, z=\pi R_2/2)$ 
is the fixed point under one projection, so, the 3-brane at that
fixed point has $N=2$ 4-dimensional supersymmetry or 8 real supercharges.
Considering the zero modes, the 4-dimensional $N=4$ $SU(5)$ gauge symmetry
is broken
down to the $N=1$ Standard Model gauge symmetry, i. e., 
$SU(5)/P^{y'z'}=SU(3)\times SU(2)\times U(1)$. If we put the Standard
Model
fermions on the observable 3-brane at $(0, 0)$ or $(0, \pi R_2)$, 
the gauge symmetry (including the KK modes) is $SU(5)$. In order to give
the Higgs
triplet large mass to avoid the proton decay, we will have 
to put the two Higgs 5-plets on the 4-brane at $y=0$ or $z=0$ to give
them masses about $1/R_1$ or $1/R_2$, which is at the order of
 unification scale. Moreover, if we put the Standard Model
fermions on the observable 3-brane at $(\pi R_1/2, \pi R_2/2)$, 
the gauge symmetry (including the KK modes) is $N=2$ 
 $SU(3)\times SU(2)\times U(1)$
 (for zero modes, $N=1$), we can put the two Higgs doublets
on the observable 3-brane or put the two Higgs 5-plets
 on the 4-brane at $y=0$ or $z=0$. The GUT scale can be at TeV scale if
we considered the SM fermions do not form the full $10^i+\bar 5^i$ $SU(5)$
multiplets
where $i$ is the generation index because there are no proton decay
problem at all.
The wrong prediction of the first and second generation mass ratio in the
usual 4-dimensional $SU(5)$ is avoided, and we will
have rich physics at TeV scale. However, 
we lose the charge quantization in the usual 4-dimensional $SU(5)$ models.
If we considered the SM fermions form the full $10^i+\bar 5^i$ $SU(5)$
multiplets,
the KK modes of $\Sigma_5^{\hat a}$ and $\Sigma_6^{\hat a}$ can induce
the proton decay by box diagram, so, the GUT scale can not be much
lower than $10^{16}$ GeV. Furthermore, if we put the
SM fermions and two Higgs 5-plets on the observable 4-branes at $y=0$
or $z=0$, we will double $10^i+\bar 5^i$ $SU(5)$ hypermultiplets, and
each generation comes from two $10 +\bar 5$ hypermultiplets due to the
choice of projection. We still have the charge quantization and
 do not have the wrong prediction of the first and second generation mass
ratio in the
usual 4-dimensional SU(5). The proton decay through dimension 5 operators
can be avoided by suitable choice of $U(1)_R$ charge for the particles.
 And the proton
decay through dimension 6 operators by exchange the gauge fields or
scalar Higgs are absent at tree level, but, we do have the box diagram for
proton decay, for example, exchange $X$ and $Y$ in the mean time, 
therefore, the GUT scale can not be much lower than $10^{16}$ GeV.

\subsection{$SU(5)$ on $M^4\times S^1/(Z_2)^2\times S^1/Z_2$}

\renewcommand{\arraystretch}{1.4}
\begin{table}[t]
\caption{Parity assignment and masses ($n\ge 0, m \ge 0$) for the vector
multiplet in 
 the $SU(5)$ model on $S^1/(Z_2)^2\times S^1/Z_2$.
\label{tab:SUV1}}
\vspace{0.4cm}
\begin{center}
\begin{tabular}{|c|c|c|}
\hline        
$(P^y, P^{y'}, P^z)$ & field & mass\\ 
\hline
$(+, +, +)$ &  $V^a_{\mu}$ & $\sqrt {(2n)^2/R_1^2+ (m)^2/R_2^2}$ \\
\hline
$(+, -, +)$ &  $V^{\hat{a}}_{\mu}$ &  $\sqrt {(2n+1)^2/R_1^2+(m)^2/R_2^2}$
\\
\hline
$(-, -, +)$ &  $\Sigma_5^a$ & $\sqrt {(2n+2)^2/R_1^2+ (m)^2/R_2^2}$\\
\hline
$(-, +, +)$ &  $\Sigma_5^{\hat{a}}$ & $\sqrt {(2n+1)^2/R_1^2+
(m)^2/R_2^2}$\\
\hline
$(+, +, -)$ &  $\Sigma_6^a$ & $\sqrt {(2n)^2/R_1^2+ (m+1)^2/R_2^2}$\\
\hline
$(+, -, -)$ &  $\Sigma_6^{\hat{a}}$ & $\sqrt {(2n+1)^2/R_1^2+
(m+1)^2/R_2^2}$\\
\hline
$(-, -, -)$ &  $\Phi^a$ & $\sqrt {(2n+2)^2/R_1^2+ (m+1)^2/R_2^2}$\\
\hline
$(-, +, -)$ &  $\Phi^{\hat{a}}$ & $\sqrt {(2n+1)^2/R_1^2+(m+1)^2/R_2^2}$
\\
\hline
\end{tabular}
\end{center}
\end{table}

\renewcommand{\arraystretch}{1.4}
\begin{table}[t]
\caption{For the model $SU(5)$ on $S^1/(Z_2)^2\times S^1/Z_2$, 
the gauge superfields, the
number of 4-dimensional supersymmetry and gauge symmetry on the 3-brane,
which
is located at the fixed point $(y=0, z=0),$ $(y=0, z=\pi R_2),$ $(y=\pi
R_1/2, z=0)$, and 
$(y=\pi R_1/2, z=\pi R_2)$, and on the 4-brane which is located at the
fixed line
$y=0$, $z=0$, $y=\pi R_1/2$, $z=\pi R_2$.
\label{tab:SUV11}}
\vspace{0.4cm}
\begin{center}
\begin{tabular}{|c|c|c|c|}
\hline        
Brane Position & Fields & SUSY & Gauge Symmetry\\ 
\hline
$(0, 0) $ &  $V^A_{\mu}$ & $N=1$ & $SU(5)$ \\
\hline
$(0, \pi R_2)$ & $V^A_{\mu}$  & N=1 & $SU(5)$ \\
\hline
$(\pi R_1/2, 0) $ & $V^a_{\mu}$, $\Sigma_5^{\hat a}$  & N=1 & $SU(3)\times
SU(2)\times U(1)$ \\
\hline
$(\pi R_1/2, \pi R_2) $ & $V^a_{\mu}$, $\Sigma_5^{\hat a}$
  & N=1 & $SU(3)\times SU(2)\times U(1)$ \\
\hline
$y=0$ & $V^A_{\mu}$, $\Sigma_6^A$  & N=2 & $SU(5)$ \\
\hline
$z= 0 $ & $V^A_{\mu}$, $\Sigma_5^A$  & N=2 & $SU(5)$ \\
\hline
$y=\pi R_1/2 $ & $V^a_{\mu}$, $\Sigma_5^{\hat a}$, $\Sigma_6^{a} $,
$\Phi^{\hat a}$
  & N=2 & $SU(3)\times SU(2)\times U(1)$ \\
\hline
$z=\pi R_2 $ & $V^A_{\mu}$, $\Sigma_5^{A}$ & N=2 & $SU(5)$ \\
\hline
\end{tabular}
\end{center}
\end{table}

Second, we will consider the $T^2/(Z_2)^3$ obtained by 
$T^2$ moduloing the equivalent classes: $y \sim -y$, $ z \sim -z$,
$y' \sim -y'$. The fixed points are $(y=0, z=0)$, $(y=0, z=\pi R_2)$,
$(y=\pi R_1/2, z=0)$ and $(y=\pi R_1/2, z=\pi R_2)$,
and the fixed lines are $y=0$, $z=0$, $y=\pi R_1/2$ and
$z=\pi R_2$. The extra space orbifold is a rectangle.

For a generic bulk field $\phi(x^{\mu}, y, z)$,
we can define three parity operators $P^y$, $P^{y'}$ and $P^z$,
respectively.
Denoting the field with ($P^y$, $P^{y'}$, $P^{z}$)=($\pm, \pm, \pm$) by 
$\phi_{\pm \pm \pm}$, we can obtain the KK mode expansions, which are
given in 
Appendix B.

We choose the following matrix representations for the parity operators
$P^y$, $P^{y'}$ and $P^z$, 
which are expressed in the adjoint representation of SU(5)
\begin{equation}
P^y={\rm diag}(+1, +1, +1, +1, +1)
~,~ P^z={\rm diag}(+1, +1, +1, +1, +1)~,~\,
\end{equation}
\begin{equation}
P^{y'}={\rm diag}(-1, -1, -1, +1, +1)
 ~.~\,
\end{equation}

Similarly, under $P^{y'}$ parity,
the $SU(5)$ gauge generators $T^A$ where A=1, 2, ..., 24 for SU(5)
are separated into two sets: $T^a$ are the gauge generators for
the Standard Model gauge group, and $T^{\hat a}$ are the other broken
gauge generators 
\begin{equation}
P^y~T^a~(P^y)^{-1}= T^a ~,~ P^y~T^{\hat a}~(P^y)^{-1}= T^{\hat a}
~,~\,
\end{equation}
\begin{equation}
P^z~T^a~(P^z)^{-1}= T^a ~,~ P^z~T^{\hat a}~(P^z)^{-1}= T^{\hat a}
~,~\,
\end{equation}
\begin{equation}
P^{y'}~T^a~(P^{y'})^{-1}= T^a ~,~ P^{y'}~T^{\hat a}
~(P^{y'})^{-1}= - T^{\hat a}
~.~\,
\end{equation}

The particle spectra are given in
 Table 3. We also present the gauge superfields, the number of
4-dimensional
supersymmetry, and gauge group on the 3-brane or 4-brane in Table 4.
Including the KK modes,
the 3-brane and 4-brane preserve $N=1$ and $N=2$ supersymmetry,
respectively. 
Considering only the zero modes, the bulk 4-dimensional
 $N=4$ $SU(5)$ gauge symmetry is broken
down to the $N=1$ SM gauge symmetry. 
The phenomenology discussions are similar to those in above
subsection, so we will not repeat them here.

\section{$N=2$ Supersymmetric $SU(5), SU(6)$ and $SO(10)$ on $M^4 \times
T^2/(Z_2)^4$}
In this section, we would like to discuss $N=2$ supersymmetric
$SU(5)$, $SU(6)$ and $SO(10)$ models on the space-time $M^4 \times
T^2/(Z_2)^4$
 without gauge-Higgs unification. We require that: (1) 
there are no zero modes for the
chiral multiplets $\Sigma_5$, $\Sigma_6$ and $\Phi$; (2)
 for the zero modes, we only have $N=1$ supersymmetric
$SU(3)\times SU(2)\times U(1)^{n-3}$
model where $n$ is the rank of the GUT group.

The orbifold $T^2/(Z_2)^4$ is
 obtained by $T^2$ moduloing the equivalence
classes: $y \sim -y$, $ z \sim -z$,
$y' \sim -y'$ and $z' \sim -z'$. The four fixed points are
$(y=0, z=0),$ $(y=0, z=\pi R_2/2),$ $(y=\pi R_1/2, z=0)$, and 
$(y=\pi R_1/2, z=\pi R_2/2)$, and the fixed lines
are $y=0$, $z=0$, $y=\pi R_1/2$ and $z=\pi R_2/2$.
Including the KK modes, the
 3-branes at the fixed points preserve $N=1$ supersymmetry,
and the 4-branes at the fixed lines preserve $N=2$
supersymmetry because the fixed points and fixed lines are fixed
under two and one projection, respectively.

We will choose the unit matrix representations for $P^y$ and $P^z$ in the
adjoint representation of GUT gauge group. So, considering the zero modes,
under $P^y$ projection, we can break the 4-dimensional $N=4$ supersymmetry
to
$N=2$ supersymmetry with $(V, \Sigma_6)$ forming a vector multiplet and
$(\Sigma_5, \Phi)$ forming a hypermultiplet, and we can break the
4-dimensional $N=2$ supersymmetry to $N=1$ supersymmetry further by
$P^z$ projection. 

For a generic bulk field $\phi(x^{\mu}, y, z)$,
we can define four parity operators $P^y$, $P^{y'}$, $P^{z}$ and $P^{z'}$,
respectively.
Denoting the field with ($P^y$, $P^{y'}$, $P^{z}$, $P^{ z'}$)=($\pm, \pm,
\pm, \pm$) by 
$\phi_{\pm \pm \pm \pm}$, we obtain the KK mode expansions, which are
given in 
Appendix C.

\subsection{$SU(5)$ Model}

\renewcommand{\arraystretch}{1.4}
\begin{table}[t]
\caption{Parity assignment and masses ($n\ge 0, m \ge 0$) for 
the vector multiplet in the $SU(5)$ model on $T^2/(Z_2)^4$.
\label{tab:SUV1}}
\vspace{0.4cm}
\begin{center}
\begin{tabular}{|c|c|c|}
\hline        
$(P^y, P^{y'}, P^z, P^{z'})$ & field & mass\\ 
\hline
$(+, +, +, +)$ &  $V^a_{\mu}$ & $\sqrt {(2n)^2/R_1^2+ (2m)^2/R_2^2}$ \\
\hline
$(+,-, +, -)$ &  $V^{\hat{a}}_{\mu}$ & $\sqrt
{(2n+1)^2/R_1^2+(2m+1)^2/R_2^2}$ \\
\hline
$(-, -, +, +)$ &  $\Sigma_5^a$ & $\sqrt {(2n+2)^2/R_1^2+ (2m)^2/R_2^2}$ \\
\hline
$(-, +, +, -)$ &  $\Sigma_5^{\hat{a}}$ & $\sqrt {(2n+1)^2/R_1^2+
(2m+1)^2/R_2^2}$ \\
\hline
$(+, +, -, -)$ &  $\Sigma_6^a$ & $\sqrt {(2n)^2/R_1^2+ (2m+2)^2/R_2^2}$\\
\hline
$(+, -, -,  +)$ &  $\Sigma_6^{\hat{a}}$ & $\sqrt {(2n+1)^2/R_1^2+
(2m+1)^2/R_2^2}$ \\
\hline
$(-, -, -, -)$ &  $\Phi^a$ & $\sqrt {(2n+2)^2/R_1^2+ (2m+2)^2/R_2^2}$\\
\hline
$(-, +, -, +)$ &  $\Phi^{\hat{a}}$ & $\sqrt
{(2n+1)^2/R_1^2+(2m+1)^2/R_2^2}$\\
\hline
\end{tabular}
\end{center}
\end{table}

\renewcommand{\arraystretch}{1.4}
\begin{table}[t]
\caption{For the model $SU(5)$ on $T^2/(Z_2)^4$,
 the gauge superfields, the number of
4-dimensional supersymmetry and gauge symmetry on the 3-brane which
is located at the fixed point $(y=0, z=0),$ $(y=0, z=\pi R_2/2),$ $(y=\pi
R_1/2, z=0)$, and 
$(y=\pi R_1/2, z=\pi R_2/2)$, or on the 4-brane which is located at the
fixed line
$y=0$, $z=0$, $y=\pi R_1/2$, $z=\pi R_2/2$.
\label{tab:SUV11}}
\vspace{0.4cm}
\begin{center}
\begin{tabular}{|c|c|c|c|}
\hline        
Brane Position & Fields & SUSY & Gauge Symmetry\\ 
\hline
$(0, 0) $ &  $V^A_{\mu}$ & $N=1$ & $SU(5)$ \\
\hline
$(0, \pi R_2/2)$ & $V^a_{\mu}$, $\Sigma_6^{\hat a}$  & N=1 & $SU(3)\times
SU(2)\times U(1)$ \\
\hline
$(\pi R_1/2, 0) $ & $V^a_{\mu}$, $\Sigma_5^{\hat a}$  & N=1 & $SU(3)\times
SU(2)\times U(1)$ \\
\hline
$(\pi R_1/2, \pi R_2/2) $ & $V^a_{\mu}$, $\Phi^{\hat a}$
  & N=1 & $SU(3)\times SU(2)\times U(1)$ \\
\hline
$y=0$ & $V^A_{\mu}$, $\Sigma_6^A$  & N=2 & $SU(5)$ \\
\hline
$z= 0 $ & $V^A_{\mu}$, $\Sigma_5^A$  & N=2 & $SU(5)$ \\
\hline
$y=\pi R_1/2 $ & $V^a_{\mu}$, $\Sigma_5^{\hat a}$, $\Sigma_6^{a} $,
$\Phi^{\hat a}$
  & N=2 & $SU(3)\times SU(2) \times U(1)$ \\
\hline
$z=\pi R_2/2 $ & $V^a_{\mu}$, $\Sigma_5^{a}$, $\Sigma_6^{\hat a} $,
$\Phi^{\hat a}$
  & N=2 & $SU(3)\times SU(2) \times U(1)$ \\
\hline
\end{tabular}
\end{center}
\end{table}

First, we will discuss the $SU(5)$ model because it is the simplest model,
 and can be generalized to other GUT models easily by our procedure.
We choose the following matrix representations 
for parity operators $P^y$, $P^z$, $P^{y'}$ and $P^{z'}$,
which are expressed in the adjoint representation of SU(5)\footnote{If one
chose
$ P^{z'}={\rm diag}(+1, +1, +1, +1, +1)$, then, the discussions are
similar to those
in subsection 3.2.}
\begin{equation}
P^y={\rm diag}(+1, +1, +1, +1, +1)
~,~ P^z={\rm diag}(+1, +1, +1, +1, +1)~,~\,
\end{equation}
\begin{equation}
P^{y'}={\rm diag}(-1, -1, -1, +1, +1)
~,~ P^{z'}={\rm diag}(-1, -1, -1, +1, +1)
 ~.~\,
\end{equation}
Under $P^{y'}$ or $P^{z'}$ parity,
the $SU(5)$ gauge generators $T^A$ where A=1, 2, ..., 24 for SU(5)
are separated into two sets: $T^a$ are the gauge generators for
the Standard Model gauge group, and $T^{\hat a}$ are the other broken
gauge generators 
\begin{equation}
P^y~T^a~(P^y)^{-1}= T^a ~,~ P^y~T^{\hat a}~(P^y)^{-1}= T^{\hat a}
~,~\,
\end{equation}
\begin{equation}
P^z~T^a~(P^z)^{-1}= T^a ~,~ P^z~T^{\hat a}~(P^z)^{-1}= T^{\hat a}
~,~\,
\end{equation}
\begin{equation}
P^{y'}~T^a~(P^{y'})^{-1}= T^a ~,~ P^{y'}~T^{\hat a}~(P^{y'})^{-1}= -
T^{\hat a}
~,~\,
\end{equation}
\begin{equation}
P^{z'}~T^a~(P^{z'})^{-1}= T^a ~,~ P^{z'}~T^{\hat a}~(P^{z'})^{-1}= -
T^{\hat a}
~.~\,
\end{equation}

The particle spectra are given in Table 5. And we present the gauge 
superfields, the number of 4-dimensional
supersymmetry, and gauge group on the 3-brane or 4-brane in Table 6.
Including the KK modes,
the 3-brane and 4-brane preserve $N=1$ and $N=2$ supersymmetry,
respectively. And the gauge group on the 3-brane or 4-brane can be
$SU(5)$ or $SU(3)\times SU(2)\times U(1)$.
Considering the zero modes, the bulk 4-dimensional
$N=4$ $SU(5)$ gauge symmetry is broken
down to the $N=1$ SM gauge symmetry. 
The phenomenology discussions are similar to those in last section.

\subsection{$SU(6)$ Model}

Second, we will discuss the $SU(6)$ model.
We need to choose the matrix representations 
for parity operators $P^y$, $P^z$, $P^{y'}$ and $P^{z'}$,
 which are expressed in the adjoint representation of SU(6).
Because $SU(6) \supset SU(5)\times U(1); SU(4)\times SU(2)\times U(1);
SU(3)\times SU(3)\times U(1)$, we obtain that, in general, 
$P^{y'}$ and $P^{z'}$ just need to be any two different representations
from
these three representations: ${\rm diag}(+1, +1, +1, +1, +1, -1)$,
${\rm diag}(-1, -1, -1, +1, +1, -1)$, and ${\rm diag}(-1, -1, -1, +1, +1,
+1)$.
So, the matrix representations for $P^y$, $P^z$, $P^{y'}$ and $P^{z'}$ 
are~\footnote{For $SU(6)$ model in this subsection, and $SO(10)$
model in the next subsection, one can interchange matrix representations
$P^{y'}$ and $P^{z'}$, i. e.,
$P^{y'} \longleftrightarrow P^{z'}$,  and
the discussions are similar.}
\begin{equation}
P^y={\rm diag}(+1, +1, +1, +1, +1, +1)
~,~ P^z={\rm diag}(+1, +1, +1, +1, +1, +1)~,~\,
\end{equation}
\begin{equation}
P^{y'}={\rm diag}(+1, +1, +1, +1, +1, -1)
~,~ P^{z'}={\rm diag}(-1, -1, -1, +1, +1, -1)
 ~,~\,
\end{equation}
or 
\begin{equation}
P^{y'}={\rm diag}(+1, +1, +1, +1, +1, -1)
~,~ P^{z'}={\rm diag}(-1, -1, -1, +1, +1, +1)
 ~,~\,
\end{equation}
or
\begin{equation}
P^{y'}={\rm diag}(-1, -1, -1, +1, +1, +1)
~,~ P^{z'}={\rm diag}(-1, -1, -1, +1, +1, -1)
 ~.~\,
\end{equation}

And we would like to point out that
\begin{equation}
SU(6)/\{{\rm diag}(+1, +1, +1, +1, +1, -1)\} \approx
SU(5)\times U(1) ~,~\,
\end{equation}
\begin{equation}
SU(6)/\{{\rm diag}(-1, -1, -1, +1, +1, -1)\} \approx
SU(4)\times SU(2) \times U(1) ~,~\,
\end{equation}
\begin{equation}
SU(6)/\{{\rm diag}(-1, -1, -1, +1, +1, +1)\} \approx
SU(3)\times SU(3) \times U(1) ~.~\,
\end{equation}

To be general, we will not specify the matrix representations
for parity operators $P^{y'}$ and $P^{z'}$ in the following
discussion. And we just use $P^y$, $P^{y'}$, $P^z$, $P^{z'}$,
 $G/P^{y'}$ and $G/P^{z'}$, where $G$ can be
$SU(6)$ in this subsection and $SO(10)$ in the next subsection.

Under $P^{y'}$ and $P^{z'}$ parities,
the $G=SU(6)$ gauge generators $T^A$ where A=1, 2, ..., 35 for SU(6)
are separated into four sets: $T^{a, b}$ are the gauge generators for
$SU(3)\times SU(2) \times U(1) \times U(1)$ gauge symmetry, $T^{ a, \hat
b}$,
$T^{\hat a, b}$, and $T^{\hat a, \hat b}$
 are the other broken gauge generators that
belong to $\{G/P^{y'} \cap \{{\rm coset}~ G/P^{z'}\}\}$,
$\{\{{\rm coset}~ G/P^{y'}\} \cap  G/P^{z'}\}$, 
and $\{\{{\rm coset}~ G/P^{y'}\} \cap \{{\rm coset}~ G/P^{z'}\}\}$,
respectively.
Therefore,
under $P^{y}$, $P^{z}$, $P^{y'}$ and $P^{z'}$, the gauge generators
transform as
\begin{equation}
P^y~T^{A, B}~(P^y)^{-1}= T^{A, B} ~,~ P^z~T^{A, B}~(P^z)^{-1}= T^{A, B} 
~,~\,
\end{equation}
\begin{equation}
P^{y'}~T^{a, B}~(P^{y'})^{-1}= T^{a, B} ~,~ 
P^{y'}~T^{\hat a, B}~(P^{y'})^{-1}= - T^{\hat a, B}
~,~\,
\end{equation}
\begin{equation}
P^{z'}~T^{A, b}~(P^{z'})^{-1}= T^{A, b} ~,~ P^{z'}~T^{A, \hat
b}~(P^{z'})^{-1}= - T^{A, \hat b}
~.~\,
\end{equation}

\renewcommand{\arraystretch}{1.4}
\begin{table}[t]
\caption{Parity assignment and masses ($n\ge 0, m \ge 0$) for the
vector multiplet in the $SU(6)$ or $SO(10)$ models on $T^2/(Z_2)^4$.
\label{tab:SUV1}}
\vspace{0.4cm}
\begin{center}
\begin{tabular}{|c|c|c|}
\hline        
$(P^y, P^{y'}, P^z, P^{z'})$ & field & mass\\ 
\hline
$(+, +, +, +)$ &  $V^{a, b}_{\mu}$ & $\sqrt {(2n)^2/R_1^2+ (2m)^2/R_2^2}$
\\
\hline
$(+, +, +, -)$ &  $V^{a, {\hat b}}_{\mu}$ & $\sqrt {(2n)^2/R_1^2+
(2m+1)^2/R_2^2}$ \\
\hline
$(+, -, +, +)$ &  $V^{{\hat a}, b}_{\mu}$ & $\sqrt {(2n+1)^2/R_1^2+
(2m)^2/R_2^2}$ \\
\hline
$(+,-, +, -)$ &  $V^{\hat{a}, \hat{b}}_{\mu}$ & $\sqrt
{(2n+1)^2/R_1^2+(2m+1)^2/R_2^2}$ \\
\hline
$(-, -, +, +)$ &  $\Sigma_5^{a, b}$ & $\sqrt {(2n+2)^2/R_1^2+
(2m)^2/R_2^2}$ \\
\hline
$(-, -, +, -)$ &  $\Sigma_5^{a, {\hat b}}$ & $\sqrt {(2n+2)^2/R_1^2+
(2m+1)^2/R_2^2}$ \\
\hline
$(-, +, +, +)$ &  $\Sigma_5^{{\hat a}, b}$ & $\sqrt {(2n+1)^2/R_1^2+
(2m)^2/R_2^2}$ \\
\hline
$(-, +, +, -)$ &  $\Sigma_5^{\hat{a}, \hat{b}}$ & $\sqrt {(2n+1)^2/R_1^2+
(2m+1)^2/R_2^2}$ \\
\hline
$(+, +, -, -)$ &  $\Sigma_6^{a, b}$ & $\sqrt {(2n)^2/R_1^2+
(2m+2)^2/R_2^2}$\\
\hline
$(+, +, -, +)$ &  $\Sigma_6^{a, {\hat b}}$ & $\sqrt {(2n)^2/R_1^2+
(2m+1)^2/R_2^2}$\\
\hline
$(+, -, -, -)$ &  $\Sigma_6^{{\hat a}, b}$ & $\sqrt {(2n+1)^2/R_1^2+
(2m+2)^2/R_2^2}$\\
\hline
$(+, -, -,  +)$ &  $\Sigma_6^{\hat{a}, {\hat b}}$ & $\sqrt
{(2n+1)^2/R_1^2+ (2m+1)^2/R_2^2}$ \\
\hline
$(-, -, -, -)$ &  $\Phi^{a, b}$ & $\sqrt {(2n+2)^2/R_1^2+
(2m+2)^2/R_2^2}$\\
\hline
$(-, -, -, +)$ &  $\Phi^{a, {\hat b}}$ & $\sqrt {(2n+2)^2/R_1^2+
(2m+1)^2/R_2^2}$\\
\hline
$(-, +, -, -)$ &  $\Phi^{{\hat a}, b}$ & $\sqrt {(2n+1)^2/R_1^2+
(2m+2)^2/R_2^2}$\\
\hline
$(-, +, -, +)$ &  $\Phi^{\hat{a}, {\hat b}}$ & $\sqrt
{(2n+1)^2/R_1^2+(2m+1)^2/R_2^2}$\\
\hline
\end{tabular}
\end{center}
\end{table}

\renewcommand{\arraystretch}{1.4}
\begin{table}[t]
\caption{For the model $G=SU(6)$ or $G=SO(10)$ on $T^2/(Z_2)^4$, the gauge
superfields, the number of 4-dimensional supersymmetry and gauge symmetry
on the 3-brane, which
is located at the fixed point $(y=0, z=0),$ $(y=0, z=\pi R_2/2),$ $(y=\pi
R_1/2, z=0)$, and 
$(y=\pi R_1/2, z=\pi R_2/2)$, or on the 4-brane which is located at the
fixed line
$y=0$, $z=0$, $y=\pi R_1/2$, $z=\pi R_2/2$.
\label{tab:SUV11}}
\vspace{0.4cm}
\begin{center}
\begin{tabular}{|c|c|c|c|}
\hline        
Brane Position & fields & SUSY & Gauge Symmetry\\ 
\hline
$(0, 0) $ &  $V^{A,B}_{\mu}$ & $N=1$ & $G$ \\
\hline
$(0, \pi R_2/2)$ & $V^{A,b}_{\mu}$, $\Sigma_6^{A, \hat b}$  & N=1 &
$G/P^{z'}$ \\
\hline
$(\pi R_1/2, 0) $ & $V^{a,B}_{\mu}$, $\Sigma_5^{\hat a, B}$  & N=1 &
$G/P^{y'} $ \\
\hline
$(\pi R_1/2, \pi R_2/2) $ & $V^{a,b}_{\mu}$, $\Sigma_5^{\hat a, b}$,
 $\Sigma_6^{a, \hat b}$, $\Phi^{\hat a, \hat b}$
  & N=1 & $SU(3)\times SU(2)\times U(1) \times U(1)$ \\
\hline
$y=0$ & $V^{A,B}_{\mu}$, $\Sigma_6^{A,B}$  & N=2 & $G$ \\
\hline
$z= 0 $ & $V^{A,B}_{\mu}$, $\Sigma_5^{A,B}$  & N=2 & $G$ \\
\hline
$y=\pi R_1/2 $ & $V^{a,B}_{\mu}$, $\Sigma_5^{\hat a, B}$, $\Sigma_6^{a, B}
$, $\Phi^{\hat a, B}$
  & N=2 & $G/P^{y'}$ \\
\hline
$z=\pi R_2/2 $ & $V^{A,b}_{\mu}$, $\Sigma_5^{A,b}$, $\Sigma_6^{A, \hat b}
$, $\Phi^{A, \hat b}$
  & N=2 & $G/P^{z'}$ \\
\hline
\end{tabular}
\end{center}
\end{table}

The particle spectra are given in Table 7, and the gauge superfields,
the number of 4-dimensional supersymmetry and  the gauge group on the
3-brane or 4-brane are given in Table 8. These two tables are very
general, because we do not specify the gauge group. And the $SU(5)$ model
in previous subsection is a special case where $T^{a \hat b}$ and $T^{\hat
a b}$
are empty sets. For the zero modes, we only have 4-dimensional $N=1$  
$SU(3)\times SU(2)\times U(1)\times U(1)$ gauge symmetry.
From Table 8, we obtain that including the KK modes,
the 3-brane and 4-brane preserve $N=1$ and $N=2$ supersymmetry,
respectively.
The gauge group on the 3-brane can be 
$G=SU(6)$, or $G/P^{z'}$, or $G/P^{y'}$, or $SU(3)\times SU(2)\times
U(1)\times U(1)$,
where $G/P^{z'}$ and $G/P^{y'}$ can be any two different gauge groups
of $SU(5)\times U(1)$, $SU(4)\times SU(2) \times U(1)$, and
$SU(3) \times SU(3) \times U(1)$. And the gauge group on the
4-brane can be $G=SU(6)$, or $G/P^{y'}$ or $G/P^{z'}$. If we
want to break the $SU(6)$ gauge group down to
the $SU(3)\times SU(2)\times U(1)\times U(1)$ completely, we will have to
put the Standard Model fermions on the 3-brane at $(\pi R_1/2, \pi
R_2/2)$.
We can put two Higgs doublets
on the observable 3-brane or put two Higgs multiplets 
 on the 4-brane at $y=\pi R_1/2$ or $z=\pi R_2/2$. 
The GUT scale can be at TeV scale if
we considered the SM fermions are just charged under
$SU(3)\times SU(2)\times U(1)\times U(1)$ gauge symmetry,
because there are no unification realization of
the SM particles, i.e., the SM fermions do not
form the multiplets under the unification gauge group, so, 
there are no proton decay problem at all.
The wrong prediction of the first and second generation mass ratio in the
usual 4-dimensional $SU(5)$ is avoided, and we will
have rich physics at TeV scale.
However,  we can not explain  the charge quantization.

We can also put the SM model fermions and Higgs on the other 3-brane
at other fixed point or on the 4-brane at fixed line.
 However, the Standard Model fermions must form the full
multiplets under the gauge group 
$SU(5)\times U(1)$, or $SU(4)\times SU(2) \times U(1)$, or
$SU(3) \times SU(3) \times U(1)$, or $SU(6)$. In addition,
if we put the SM fermions on the 3-brane at $(0,0)$, or $(0, \pi R_2/2)$,
or $(\pi R_1/2, 0)$,
in order to give the GUT scale ($1/R_1$
or $1/R_2$) masses to the Higgs components, which do
not belong to $SU(2)$ doublets, from the projections,
 we may need to put the Higgs on the 4-brane. 
And if we put the SM fermions on the 4-brane, 
we will double the generations due to the projections, i. e.,
each generation in the SM comes from the zero modes of 
two generations. 
Moreover, although at tree level the proton decay may
be absent, the GUT scale can not be much lower than
$10^{16}$ GeV because the proton may decay through box diagrams.

Furthermore, in order to break the extra $U(1)$ symmetry, we have to
introduce the chiral multiplets that are singlets under
the SM gauge symmetry, and use Higgs mechanism.
 And if we considered the chiral model on the observable brane, we will
have to introduce exotic particles due to the anomaly cancellation.

\subsection{$SO(10)$ Model}

Third, we would like to discuss the $SO(10)$ model.
We choose the following matrix representations for the parity operators
 $P^y$, $P^z$, $P^{y'}$, and $P^{z'}$,
which are expressed in the adjoint representation of $SO(10)$

\begin{equation}
P^y={\rm diag}(+\sigma_0, +\sigma_0, +\sigma_0, +\sigma_0, +\sigma_0)
~,~\,
\end{equation}
\begin{equation}
P^z={\rm diag}(+\sigma_0, +\sigma_0, +\sigma_0, +\sigma_0, +\sigma_0)~,~\,
\end{equation}
\begin{equation}
P^{y'}={\rm diag}(\sigma_2, \sigma_2, \sigma_2, \sigma_2, \sigma_2)
~,~\,
\end{equation}
\begin{equation}
P^{z'}={\rm diag}(-\sigma_0, -\sigma_0, -\sigma_0, +\sigma_0, +\sigma_0)
 ~,~\,
\end{equation}
where $\sigma_0$ is the $2\times 2$ unit matrix and  $\sigma_2$ is the 
Pauli matrix. 

Under $P^{y'}$ projection, the gauge group $SO(10)$ is
broken down to $SO(10)/P^{y'}=SU(5)\times U(1)$,
and under $P^{z'}$ projection, the gauge group is broken down to the
$SO(10)/P^{z'}=SO(6)\times SO(4)$ which is isomorphic to $SU(4)\times
SU(2)\times SU(2)$. 
Although $SO(10) \supset SU(5)\times U(1); SU(4)\times SU(2)\times SU(2);
SO(8)\times U(1)$, if we want to break $SO(10)$ down to 
$SU(3)\times SU(2) \times U(1) \times U(1)$ completely, we can not choose
either
\begin{equation}
P^{y'}={\rm diag}(-\sigma_0, -\sigma_0, -\sigma_0, -\sigma_0, +\sigma_0)
 ~,~\,
\end{equation}
or
\begin{equation}
P^{z'}={\rm diag}(-\sigma_0, -\sigma_0, -\sigma_0, -\sigma_0, +\sigma_0)
 ~,~\,
\end{equation}
which will break $SO(10)$ down to $SO(8)\times U(1)$.

Under $P^{y'}$ and $P^{z'}$ parities,
the $SO(10)$ gauge generators $T^A$ where A=1, 2, ..., 45 for $SO(10)$
are separated into four sets: $T^{a, b}$ are the gauge generators for
$SU(3)\times SU(2)\times U(1)\times U(1)$ gauge symmetry, $T^{a, \hat b}$,
$T^{\hat a, b}$, and $T^{\hat a, \hat b}$
 are the other broken
gauge generators in $\{G/P^{y'} \cap \{{\rm coset}~ G/P^{z'}\}\}$,
$\{\{{\rm coset}~ G/P^{y'}\} \cap  G/P^{z'}\}$, 
and $\{\{{\rm coset}~ G/P^{y'}\} \cap \{{\rm coset}~ G/P^{z'}\}\}$,
respectively.
So, under $P^{y}$, $P^{z}$, $P^{y'}$ and $P^{z'}$, the gauge generators
transform as
\begin{eqnarray}
P^y~T^{A, B}~(P^y)^{-1}= T^{A, B} ~,~ P^z~T^{A, B}~(P^z)^{-1}= T^{A, B} 
~,~\,
\end{eqnarray}
\begin{eqnarray}
P^{y'}~T^{a, B}~(P^{y'})^{-1}= T^{a, B} ~,~ 
P^{y'}~T^{\hat a, B}~(P^{y'})^{-1}= - T^{\hat a, B}
~,~\,
\end{eqnarray}
\begin{eqnarray}
P^{z'}~T^{A, b}~(P^{z'})^{-1}= T^{A, b} ~,~ P^{z'}~T^{A, \hat
b}~(P^{z'})^{-1}= - T^{A, \hat b}
~.~\,
\end{eqnarray}

The particle spectra are given in Table 7. And the gauge superfields,
the number of 4-dimensional supersymmetry, and the gauge group on the
3-brane or 4-brane are given in Table 8.
For the zero modes, we have 4-dimensional $N=1$ 
$SU(3)\times SU(2)\times U(1)\times U(1)$ gauge symmetry.
From Table 8, we obtain that including the KK modes,
the 3-brane and 4-brane preserve $N=1$ and $N=2$ supersymmetry,
respectively. 
The gauge group on the 3-brane can be 
$G=SO(10)$, or $SU(5)\times U(1)$, or $SU(4)\times SU(2)\times SU(2)$,
 or $SU(3)\times SU(2)\times U(1)\times U(1)$.
 And the gauge group on the
4-brane can be $G=SU(6)$, or $SU(5)\times U(1)$, or $SU(4)\times
SU(2)\times SU(2)$.
The phenomenology discussions are similar to those in last subsection, and
we want to point out that
$SO(10)$ is a popular GUT model where one generation in the SM fits
perfectly in a 16 representation of $SO(10)$.

\section{Gauge-Higgs Unification for $SU(6)$, $SU(7)$ and $SO(12)$}
It is noticed that the gauge-Higgs might be unified at a vector multiplets
in 6-dimensional model~\cite{HNDS}. 
In this section, we would like to discuss the
$SU(N)$ and $SO(M)$ GUT models which have this property. Our basic
requirements
 are: (1) there are no zero modes for the chiral multiplets $\Sigma_5$ 
and $\Sigma_6$; 
(2) considering the zero modes,
there are only one pair of Higgs doublets because if
we had two pairs of Higgs doublets, we may have flavour changing
neutral current problem.  

The minimal model for $SU(N)$ gauge group is $SU(6)$ on the space-time
$M^4\times T^2/(Z_2)^3$. The proof is following:
we have six projection operators $P^{y}$, $P^{z}$, $P^{yz}$,
$P^{y'}$, $P^{z'}$, and $P^{y'z'}$, in order to break $SU(6)$ down to
$SU(3)\times SU(2) \times U(1) \times U(1)$, we need at least
two projection operators. The non-equivalent pair choices are
$(P^{y}, P^{z})$, $(P^{y}, P^{y'})$ and $(P^{yz}, P^{y' z'})$. And
it is not difficult for one to show that all of these choice do
 not satisfy
our basic requirements. Similarly, the minimal model for $SO(M)$ gauge
group
is $SO(12)$ on the space-time $M^4\times T^2/(Z_2)^4$. We also discuss the
$SU(7)$ model because they are the only models which satisfy our
requirements, and have the gauge-Higgs unification
on the extra space orbifold $T^2/(Z_2)^3$ and $T^2/(Z_2)^4$.
We also comment on 
$SU(7)$ and $SO(12)$ models on $T^2/(Z_2)^3$ which do not
satisfy our requirements.

In order to project out all the zero modes of $\Sigma_5$ and $\Sigma_6$,
we would like to choose the matrix representation of $P^{y}$
is equal to that of $P^{z}$, which is equivalent to choosing
$P^{yz}$ as an unit matrix, and $P^y$ as other matrix. And it is 
easy to see that for zero modes,
$P^{yz}$ breaks 4-dimensional $N=4$ supersymmetry to $N=2$ supersymmetry,
 and projects out all the zero modes of $\Sigma_5$ and $\Sigma_6$. 
In order to match our convention in~\cite{TJ}, 
we would like to choose the matrix representation for $P^{y}$
to be equal to that for $P^{z}$.

\subsection{$SU(6)$ on $M^4\times T^2/(Z_2)^3$}

As we know, $SU(6) \supset SU(5)\times U(1); SU(4)\times SU(2)\times U(1);
SU(3)\times SU(3)\times U(1)$, so, we have three projection pair choices
to have the gauge-Higgs unification, and
we will discuss all of them because they are the bases to consider the
gauge-Higgs unification for the other higher rank semi-simple groups.

(I) We choose the matrix representations for $P^y$, $P^z$
 and $P^{y'z'}$ as following
\begin{equation}
P^y=P^z={\rm diag}(+1, +1, +1, +1, +1, -1)
~,~\,
\end{equation}
\begin{equation}
P^{y'z'}={\rm diag}(-1, -1, -1, +1, +1, +1)~.~\,
\end{equation}

To identify the massless fields, we consider the transformation 
properties of $V$ and $\Phi$ fields under $P^y$ and 
$P^{y' z'}$, i.e., $(P^y, P^{y' z'})$
\begin{equation}
  V:\: \left( \begin{array}{ccc|cc|c}
    (+,+) & (+,+) & (+,+) & (+,-) & (+,-) & (-,-) \\ 
    (+,+) & (+,+) & (+,+) & (+,-) & (+,-) & (-,-) \\ 
    (+,+) & (+,+) & (+,+) & (+,-) & (+,-) & (-,-) \\ \hline
    (+,-) & (+,-) & (+,-) & (+,+) & (+,+) & (-,+) \\ 
    (+,-) & (+,-) & (+,-) & (+,+) & (+,+) & (-,+) \\ \hline
    (-,-) & (-,-) & (-,-) & (-,+) & (-,+) & (+,+)
  \end{array} \right),
\label{eq:V6trans}
\end{equation}
\begin{equation}
  \Phi:\: \left( \begin{array}{ccc|cc|c}
    (-,+) & (-,+) & (-,+) & (-,-) & (-,-) & (+,-) \\ 
    (-,+) & (-,+) & (-,+) & (-,-) & (-,-) & (+,-) \\ 
    (-,+) & (-,+) & (-,+) & (-,-) & (-,-) & (+,-) \\ \hline
    (-,-) & (-,-) & (-,-) & (-,+) & (-,+) & (+,+) \\ 
    (-,-) & (-,-) & (-,-) & (-,+) & (-,+) & (+,+) \\ \hline
    (+,-) & (+,-) & (+,-) & (+,+) & (+,+) & (-,+)
  \end{array} \right).
\label{eq:phi6trans}
\end{equation}

(II) We choose the matrix representations for $P^y$, $P^z$,
 and $P^{y'z'}$ as following
\begin{equation}
P^y=P^z={\rm diag}(-1, -1, -1, +1, +1, -1)
~,~\,
\end{equation}
\begin{equation}
P^{y'z'}={\rm diag}(-1, -1, -1, +1, +1, +1)~.~\,
\end{equation}

To identify the massless fields, we consider the transformation 
properties of $V$ and $\Phi$ fields under $P^y$ and 
$P^{y' z'}$, i.e., $(P^y, P^{y' z'})$
\begin{equation}
  V:\: \left( \begin{array}{ccc|cc|c}
    (+,+) & (+,+) & (+,+) & (-,-) & (-,-) & (+,-) \\ 
    (+,+) & (+,+) & (+,+) & (-,-) & (-,-) & (+,-) \\ 
    (+,+) & (+,+) & (+,+) & (-,-) & (-,-) & (+,-) \\ \hline
    (-,-) & (-,-) & (-,-) & (+,+) & (+,+) & (-,+) \\ 
    (-,-) & (-,-) & (-,-) & (+,+) & (+,+) & (-,+) \\ \hline
    (+,-) & (+,-) & (+,-) & (-,+) & (-,+) & (+,+)
  \end{array} \right),
\label{eq:V6trans}
\end{equation}
\begin{equation}
  \Phi:\: \left( \begin{array}{ccc|cc|c}
    (-,+) & (-,+) & (-,+) & (+,-) & (+,-) & (-,-) \\ 
    (-,+) & (-,+) & (-,+) & (+,-) & (+,-) & (-,-) \\ 
    (-,+) & (-,+) & (-,+) & (+,-) & (+,-) & (-,-) \\ \hline
    (+,-) & (+,-) & (+,-) & (-,+) & (-,+) & (+,+) \\ 
    (+,-) & (+,-) & (+,-) & (-,+) & (-,+) & (+,+) \\ \hline
    (-,-) & (-,-) & (-,-) & (+,+) & (+,+) & (-,+)
  \end{array} \right).
\label{eq:phi6trans}
\end{equation}

(III) We choose the matrix representations for $P^y$, $P^z$,
 and $P^{y'}$ as following
\begin{equation}
P^y=P^z={\rm diag}(+1, +1, +1, +1, +1, -1)~,~\,
\end{equation} 
\begin{equation}
P^{y'}={\rm diag}(-1, -1, -1, +1, +1, -1)~,~\,
\end{equation}
or
\begin{equation}
P^y=P^z={\rm diag}(-1, -1, -1, +1, +1, -1)
~,~\,
\end{equation} 
\begin{equation}
P^{y'}={\rm diag}(+1, +1, +1, +1, +1, -1)~.~\,
\end{equation}

To identify the massless fields, we consider the transformation 
properties of $V$ and $\Phi$ fields under $P^y$ and 
$P^{y'}$, i.e., $(P^y, P^{y'})$ by the first definition of 
$P^y$, $P^z$ and $P^{y'}$
\begin{equation}
  V:\: \left( \begin{array}{ccc|cc|c}
    (+,+) & (+,+) & (+,+) & (+,-) & (+,-) & (-,+) \\ 
    (+,+) & (+,+) & (+,+) & (+,-) & (+,-) & (-,+) \\ 
    (+,+) & (+,+) & (+,+) & (+,-) & (+,-) & (-,+) \\ \hline
    (+,-) & (+,-) & (+,-) & (+,+) & (+,+) & (-,-) \\ 
    (+,-) & (+,-) & (+,-) & (+,+) & (+,+) & (-,-) \\ \hline
    (-,+) & (-,+) & (-,+) & (-,-) & (-,-) & (+,+)
  \end{array} \right),
\label{eq:V6trans}
\end{equation}
\begin{equation}
  \Phi:\: \left( \begin{array}{ccc|cc|c}
    (-,-) & (-,-) & (-,-) & (-,+) & (-,+) & (+,-) \\ 
    (-,-) & (-,-) & (-,-) & (-,+) & (-,+) & (+,-) \\ 
    (-,-) & (-,-) & (-,-) & (-,+) & (-,+) & (+,-) \\ \hline
    (-,+) & (-,+) & (-,+) & (-,-) & (-,-) & (+,+) \\ 
    (-,+) & (-,+) & (-,+) & (-,-) & (-,-) & (+,+) \\ \hline
    (+,-) & (+,-) & (+,-) & (+,+) & (+,+) & (-,-)
  \end{array} \right).
\label{eq:phi6trans}
\end{equation}

\renewcommand{\arraystretch}{1.4}
\begin{table}[t]
\caption{For the case (I) and (II), the
 $G=SU(6)$ model with gauge-Higgs unification on 
$(S^1/Z_2\times S^1/Z_2)/Z_2$.
The gauge superfield $V_{\mu}$, the number of 4-dimensional supersymmetry
and gauge 
symmetry on the 3-brane, which
is located at the fixed point $(y=0, z=0),$ $(y=0, z=\pi R_2),$  and 
$(y=\pi R_1/2, z=\pi R_2/2)$, and on the 4-brane which is located at the
fixed line
$y=0$ and $z=0$. Notice that the matrix representation for $P^y$ is the 
same as that for $P^z$.
\label{tab:SUV11}}
\vspace{0.4cm}
\begin{center}
\begin{tabular}{|c|c|c|c|}
\hline        
Brane Position & Fields & SUSY & Gauge Symmetry\\ 
\hline
$(0, 0) $ &  $V^{a,B}_{\mu}$ & $N=1$ & $G/P^y~ {\rm or}~ G/P^z$ \\
\hline
$(0, \pi R_2)$ & $V^{a,B}_{\mu}$ & N=1 & $G/P^y ~{\rm or}~ G/P^z $ \\
\hline
$(\pi R_1/2, \pi R_2/2) $ & $V^{A,b}_{\mu}$ 
  & N=2 & $G/P^{y' z'}$ \\
\hline
$y=0$ & $V^{a,B}_{\mu}$  & N=2 & $G/P^y ~{\rm or}~ G/P^z$ \\
\hline
$z= 0 $ & $V^{a,B}_{\mu}$  & N=2 & $G/P^y ~{\rm or}~ G/P^z$ \\
\hline
\end{tabular}
\end{center}
\end{table}

\renewcommand{\arraystretch}{1.4}
\begin{table}[t]
\caption{For  the case (III), the  $G=SU(6)$ model
with gauge-Higgs unification on $S^1/(Z_2)^2\times S^1/Z_2$.
The gauge superfield $V_{\mu}$, the
number of the 4-dimensional supersymmetry and gauge 
symmetry on the 3-brane, which
is located at the fixed point $(y=0, z=0),$ $(y=0, z=\pi R_2),$ $(y=\pi
R_1/2, z=0)$, and 
$(y=\pi R_1/2, z=\pi R_2)$, and on the 4-brane which is located at the
fixed line
$y=0$, $z=0$, $y=\pi R_1/2$, $z=\pi R_2$.
 Notice that the matrix representation for $P^y$ is the 
same as that for $P^z$, and $G/\{P^y \cup P^{y'}\}=SU(3)\times SU(2)\times
U(1)\times U(1)$.
\label{tab:SUV11}}
\vspace{0.4cm}
\begin{center}
\begin{tabular}{|c|c|c|c|}
\hline        
Brane Position & Fields & SUSY & Gauge Symmetry\\ 
\hline
$(0, 0) $ &  $V^{a,B}_{\mu}$ & $N=1$ & $G/P^y ~{\rm or}~ G/P^z$ \\
\hline
$(0, \pi R_2)$ & $V^{a,B}_{\mu}$  & N=1 & $G/P^y ~{\rm or}~ G/P^z$ \\
\hline
$(\pi R_1/2, 0) $ & $V^{a,b}_{\mu}$  & N=1 &
 $G/\{P^y \cup P^{y'}\}$ \\
\hline
$(\pi R_1/2, \pi R_2) $ & $V^{a,b}_{\mu}$
  & N=1 & $G/\{P^y \cup P^{y'}\}$ \\
\hline
$y=0$ & $V^{a,B}_{\mu}$  & N=2 & $G/P^y ~{\rm or}~ G/P^z$ \\
\hline
$z= 0 $ & $V^{a,B}_{\mu}$  & N=2 & $G/P^y ~{\rm or}~ G/P^z$ \\
\hline
$y=\pi R_1/2 $ & $V^{A,b}_{\mu}$
  & N=2 & $G/P^{y'}$ \\
\hline
$z=\pi R_2 $ & $V^{a,B}_{\mu}$ & N=2 & $G/P^y ~{\rm or}~ G/P^z$ \\
\hline
\end{tabular}
\end{center}
\end{table}

For all above three cases, it is not difficult to obtain the number
of 4-dimensional
supersymmetry and gauge symmetry on the 3-brane which
is located at the fixed point or on the 4-brane which
is located at the fixed line.
Because the geometry of the extra
dimensions in case (I) and (II) is different from that
in case (III), we present the number
of 4-dimensional
supersymmetry and gauge symmetry
on the 3-brane and 4-brane for case (I) and (II) in
Table 9, and those for case (III) in Table 10.
The phenomenology discussions are similar to those in
 previous sections.

\subsection{$SU(6)$, $SU(7)$ and $SO(12)$ on $M^4\times T^2/(Z_2)^4$}

Similarly, we can discuss the gauge-Higgs unification for $SU(6)$,
$SU(7)$ and $SO(10)$ on the space-time $M^4\times T^2/(Z_2)^4$. Because
there might
exist a lot of possibilities for $SU(7)$ and $SO(12)$, we just give some
 as examples. And we would like to emphasize that $SO(12)$ is the minimal
model with gauge-Higgs unification for $SO(M)$ gauge group~\footnote{In
this
subsection, one can interchange the matrix representations for 
$P^y$, $P^{y'}$, $P^{z'}$, i.e, $P^y \longleftrightarrow P^{y'}$,
$P^y \longleftrightarrow P^{z'}$, $P^{y'} \longleftrightarrow P^{z'}$.
The discussions are similar.}.

(I) For the $SU(6)$ model, similar to the case (III) in last subsection,
we can choose following matrix representations for the
parity operators $P^y$, $P^z$, 
 $P^{y'}$ and $P^{z'}$ 
\begin{equation}
P^y=P^z={\rm diag}(+1, +1, +1, +1, +1, -1)
~,~\,
\end{equation} 
\begin{equation}
P^{y'}=P^{z'}={\rm diag}(-1, -1, -1, +1, +1, -1)~.~\,
\end{equation}
Obviously,  we have the gauge-Higgs
unification.

(II) For the $SU(7)$ model, we can use above $SU(6)$ matrix
representations
for the parity operators $P^y$, $P^z$,  $P^{y'}$ and $P^{z'}$, and just
add the
 ``$\pm$'' in the matrix. Of course, there are several possibilities, for
example, we can choose the matrix representations for 
$P^y$, $P^z$, $P^{y'}$ and $P^{z'}$ as following
\begin{equation}
P^y=P^z={\rm diag}(+1, +1, +1, +1, +1, -1, \pm1 ~{\rm or}~ \mp1)~,~\,
\end{equation}
\begin{equation}
 P^{y'}={\rm diag}(-1, -1, -1, +1, +1, -1, \pm1)~,~\,
\end{equation}
\begin{equation}
 P^{z'}={\rm diag}(-1, -1, -1, +1, +1, -1, \mp1)~,~\,
\end{equation}
or 
\begin{equation}
P^y=P^z={\rm diag}(-1, -1, -1, +1, +1, -1, \pm1 ~{\rm or}~ \mp1)~,~\,
\end{equation}
\begin{equation}
 P^{y'}={\rm diag}(+1, +1, +1, +1, +1, -1, \pm1)~,~\,
\end{equation}
\begin{equation}
 P^{z'}={\rm diag}(+1, +1, +1, +1, +1, -1, \mp1)~.~\,
\end{equation}

It is easy to check that, after projections, there are only a pair of
$SU(2)$ doublets in $\Phi$ which have zero modes, and can be identified as
two
Higgs doublets. In fact, we just need to check whether the
matrix elements $(i, 6)$ and $(6, j)$ will be projected out or not,
 where $i, j=1, 2, ..., 6$.
And the quotient groups are
\begin{equation}
SU(7)/\{{\rm diag}(+1, +1, +1, +1, +1, -1, +1)\} \approx
SU(6)\times U(1) ~,~\,
\end{equation}
\begin{equation}
SU(7)/\{{\rm diag}(+1, +1, +1, +1, +1, -1, -1)\} \approx
SU(5)\times SU(2)\times U(1) ~,~\,
\end{equation}
\begin{equation}
SU(7)/\{{\rm diag}(-1, -1, -1, +1, +1, -1, -1)\} \approx
SU(5)\times SU(2) \times U(1)  ~,~\,
\end{equation}
\begin{equation}
SU(7)/\{{\rm diag}(-1, -1, -1, +1, +1, -1, +1)\} \approx
SU(4)\times SU(3) \times U(1)  ~.~\,
\end{equation}

(III) For the $SO(12)$ model, similar to the $SU(6)$ model, 
 we choose the matrix representations for the parity operators $P^y$,
$P^z$,
 $P^{y'}$ and $P^{z'}$ as following
\begin{equation}
P^y=P^z={\rm diag}(+\sigma^2, +\sigma^2, +\sigma^2, +\sigma^2, +\sigma^2,
-\sigma^2)~.~\,
\end{equation}
\begin{equation}
 P^{y'}={\rm diag}(+\sigma^0, +\sigma^0, +\sigma^0, +\sigma^0, +\sigma^0,
-\sigma^0)~.~\,
\end{equation}
\begin{equation}
 P^{z'}={\rm diag}(-\sigma^0, -\sigma^0, -\sigma^0, +\sigma^0, +\sigma^0,
-\sigma^0)~.~\,
\end{equation}
After projections, there are only a pair of $SU(2)$
 doublets in $\Phi$ which have zero modes, and can be identified as two
Higgs doublets. And the quotient groups are
\begin{equation}
SO(12)/ {\rm diag}(+\sigma^2, +\sigma^2, +\sigma^2, +\sigma^2, +\sigma^2,
-\sigma^2)
\approx SU(6)'\times U(1)' ~,~\,
\end{equation}
\begin{equation}
SO(12)/ {\rm diag}(+\sigma^0, +\sigma^0, +\sigma^0, +\sigma^0, +\sigma^0,
-\sigma^0)
 \approx
SO(10)\times U(1) ~,~\,
\end{equation}
\begin{equation}
SO(12)/{\rm diag}(-\sigma^0, -\sigma^0, -\sigma^0, +\sigma^0, +\sigma^0,
-\sigma^0)
 \approx SO(8)\times SO(4) ~.~\,
\end{equation}
By the way, $SO(4) \sim SU(2)\times SU(2)$.

Under $P^{y}$ (similar for $P^z$), $P^{y'}$ and $P^{z'}$ parities,
the $SU(7)$ and $SO(12)$ gauge generators $T^A$ where A=1, 2, ..., 48 for
$SU(7)$
or 66 for $SO(12)$
are separated into 8 sets: $T^{a b c}$ are the gauge generators for
$SU(3)\times SU(2)\times U(1)^3$, and $T^{ a b \hat c}$, $T^{ a \hat b
c}$,
$T^{ a \hat b \hat c}$, $T^{\hat a b c}$, $T^{ \hat a b \hat c}$, $T^{
\hat a \hat b c}$,
$T^{ \hat a \hat b \hat c}$
 are the other broken
gauge generators that belong to 
$\{G/P^{y} \cap G/P^{y'} \cap \{{\rm coset}~ G/P^{z'}\}\}$,
$\{G/P^{y} \cap \{{\rm coset}~G/P^{y'}\} \cap  G/P^{z'}\}$,
$\{G/P^{y} \cap \{{\rm coset}~G/P^{y'}\} \cap \{{\rm coset}~
G/P^{z'}\}\}$,
$\{\{{\rm coset}~ G/P^{y}\} \cap G/P^{y'} \cap  G/P^{z'}\}$,
$\{\{{\rm coset}~ G/P^{y}\} \cap G/P^{y'} \cap \{{\rm coset}~
G/P^{z'}\}\}$,
$\{\{{\rm coset}~ G/P^{y}\} \cap \{{\rm coset}~G/P^{y'}\} \cap
G/P^{z'}\}$,
$\{\{{\rm coset}~ G/P^{y}\}
 \cap \{{\rm coset}~G/P^{y'}\} \cap \{{\rm coset}~ G/P^{z'}\}\}$,
Therefore,
under $P^{y}$, $P^{z}$, $P^{y'}$ and $P^{z'}$, the gauge generators
transform as
\begin{equation}
P^y~T^{a, B, C}~(P^y)^{-1}= T^{a, B, C} ~,~ P^y~T^{\hat a, B,
C}~(P^y)^{-1}= 
-T^{\hat a, B, C} 
~,~\,
\end{equation}
\begin{equation}
P^z~T^{a, B, C}~(P^z)^{-1}= T^{a, B, C} ~,~ P^z~T^{\hat a, B,
C}~(P^z)^{-1}= 
-T^{\hat a, B, C} 
~,~\,
\end{equation}
\begin{equation}
P^{y'}~T^{A, b, C}~(P^{y'})^{-1}= T^{A, b, C} ~,~ 
P^{y'}~T^{A, \hat b, C }~(P^{y'})^{-1}= - T^{A, \hat b, C}
~,~\,
\end{equation}
\begin{equation}
P^{z'}~T^{A, B, c}~(P^{z'})^{-1}= T^{A, B, c} ~,~ 
P^{z'}~T^{A, B, \hat c }~(P^{z'})^{-1}= - T^{A, B, \hat c}
~.~\,
\end{equation}

\renewcommand{\arraystretch}{1.4}
\begin{table}[t]
\caption{For the model $G=SU(7)$ or $G=SO(12)$ on $T^2/(Z_2)^4$, the gauge
superfield $V_{\mu}$, the number of the 4-dimensional supersymmetry and 
gauge symmetry on the 3-brane which
is located at the fixed point $(y=0, z=0),$ $(y=0, z=\pi R_2/2),$ $(y=\pi
R_1/2, z=0)$, and 
$(y=\pi R_1/2, z=\pi R_2/2)$, or on the 4-brane which is located at the
fixed line
$y=0$, $z=0$, $y=\pi R_1/2$, $z=\pi R_2/2$. 
\label{tab:SUV11}}
\vspace{0.4cm}
\begin{center}
\begin{tabular}{|c|c|c|c|}
\hline        
Brane Position & fields & SUSY & Gauge Symmetry\\ 
\hline
$(0, 0) $ &  $V^{a,B,C}_{\mu}$ & $N=1$ & $G/P^y$ \\
\hline
$(0, \pi R_2/2)$ & $V^{a,B,c}_{\mu}$  & N=1 & $G/\{P^{y}\cup P^{z'}\}$ \\
\hline
$(\pi R_1/2, 0) $ & $V^{a,b,C}_{\mu}$  & N=1 &  $G/\{P^{y}\cup P^{y'}\}$
\\
\hline
$(\pi R_1/2, \pi R_2/2) $ & $V^{A,b,c}_{\mu}$  & N=1 &  $G/\{P^{y'}\cup
P^{z'}\}$\\
\hline
$y=0$ &  $V^{a,B,C}_{\mu}$ & $N=2$ & $G/P^y$ \\
\hline
$z= 0 $ & $V^{a,B,C}_{\mu}$ & $N=2$ & $G/P^y$ \\
\hline
$y=\pi R_1/2 $ &  $V^{A,b,C}_{\mu}$  & N=2 &  $G/ P^{y'}$  \\
\hline
$z=\pi R_2/2 $ & $V^{A,B,c}_{\mu}$  & N=2 &  $G/ P^{z'}$  \\
\hline
\end{tabular}
\end{center}
\end{table}

With the KK modes expansions in Appendix C,
we obtain the superfield $V_{\mu}$, the number
of 4-dimensional supersymmetry and gauge symmetry on the 3-brane 
and 4-brane, which are given in Table 11.
 The phenomenological discussions are
similar to those in previous section. And the $SU(6)$ model is
a special case where $T^{A b \hat c}$ and $T^{A \hat b c}$
are empty sets. In addition, for $SU(7)$, it is easy to determine
the quotient groups $G/\{P^y \cup P^{y'}\}$, $G/\{P^y \cup P^{z'}\}$,
$G/\{P^{y'} \cup P^{z'}\}$.
For $SO(12)$, they are
\begin{equation}
G/\{P^y \cup P^{y'}\}= SU(5)\times U(1) \times U(1) ~,~\,
\end{equation}
\begin{equation}
G/\{P^y \cup P^{z'}\}= SU(4)' \times SU(2) \times U(1)\times U(1) ~,~\,
\end{equation}
\begin{equation}
G/\{P^{y'} \cup P^{z'}\}= SO(6)\times SO(4)\times U(1) ~.~\,
\end{equation}
  
\subsection{Comments on $SU(7)$ and $SO(12)$ on $M^4\times T^2/(Z_2)^3$}
If one relaxed the requirements that there are no zero modes for 
chiral multiplets $\Sigma_5$ and $\Sigma_6$, we can consider 
the $SU(7)$ and $SO(12)$ models on $M^4\times T^2/(Z_2)^3$.

If the space-time orbifold was $M^4\times S^1/(Z_2)^2 \times S^1/Z_2$,
using the matrix representations of the parity operators for $SU(7)$ and
$SO(12)$ models
in last subsection, 
we can define the matrix representations for $P^y$, $P^z$ and $P^{y'}$ by
$P^y={\rm last ~subsection}~ P^{y'}$, $P^z={\rm last ~subsection}~ P^{z'}$
and $P^{y'}={\rm last ~subsection}~ P^{y}$. So, the discussions are
similar.

Now, we consider the space-time orbifold is $M^4\times (S^1/Z_2 \times
S^1/Z_2)/Z_2$.
For the $SU(7)$ model, using the
result for $SU(6)$ model, we can define the matrix representations for
$P^y$, $P^z$ 
and $P^{y'z'}$ as
\begin{equation}
P^y={\rm diag}(+1, +1, +1, +1, +1, -1, \pm1 )
~,~\,
\end{equation}
\begin{equation}
P^z={\rm diag}(+1, +1, +1, +1, +1, -1, \mp1 )
~,~\,
\end{equation}
\begin{equation}
P^{y'z'}={\rm diag}(-1, -1, -1, +1, +1, +1, \pm1 ~{\rm or}~ \mp1)~,~\,
\end{equation}
or
\begin{equation}
P^y={\rm diag}(-1, -1, -1, +1, +1, -1, \pm1)
~,~\,
\end{equation}
\begin{equation}
P^z={\rm diag}(-1, -1, -1, +1, +1, -1, \mp1)
~,~\,
\end{equation}
\begin{equation}
P^{y'z'}={\rm diag}(-1, -1, -1, +1, +1, +1, \pm1 ~{\rm or}~ \mp1)~.~\,
\end{equation}
And it is easy to check that after projections, there are only a pair of
$SU(2)$ doublets in $\Phi$ which have zero modes, and can be identified as
two
Higgs doublets.

For $SO(12)$ model, we can define the matrix representations for $P^y$,
$P^z$ 
and $P^{y'z'}$ as
\begin{equation}
 P^{y}={\rm diag}(+\sigma^0, +\sigma^0, +\sigma^0, +\sigma^0, +\sigma^0,
-\sigma^0)~,~\,
\end{equation}
\begin{equation}
 P^{z}={\rm diag}(-\sigma^0, -\sigma^0, -\sigma^0, +\sigma^0, +\sigma^0,
-\sigma^0)~,~\,
\end{equation}
\begin{equation}
P^{y'z'}={\rm diag}(+\sigma^2, +\sigma^2, +\sigma^2, +\sigma^2, +\sigma^2,
+\sigma^2)~.~\,
\end{equation}
After projections, we have only a pair of
$SU(2)$ doublets in $\Phi$ which have zero modes, and can be identified as
two
Higgs doublets. By the way,
\begin{equation}
SO(12)/ {\rm diag}(+\sigma^2, +\sigma^2, +\sigma^2, +\sigma^2, +\sigma^2,
+\sigma^2)
\approx SU(6)\times U(1) ~.~\,
\end{equation}

It is not difficult to obtain the number
of 4-dimensional supersymmetry and gauge symmetry on the 3-brane which
is located at the fixed point or on the 4-brane which
is located at the fixed line. And the phenomenological discussions are
similar to those in the previous sections.

\section{Phenomenology and Generalization Discussion}

In this section, we will discuss some phenomenological aspect
of the GUT models on the $T^2$ orbifolds and generalize
our procedure.

The first point we want to make is that, the $Z_2$ projections do not
change the rank of the gauge group, because the generators in the Cartan
subalgebra, which is the maximal Abelian subalgebra of
group $G$, are always even under the $Z_2$ projections, 
and then, we can not project them out.
Therefore, for a rank $n$ GUT group, we can only break it
down to $SU(3)\times SU(2)\times U(1)\times U(1)^{n-4}$.
In order to break the extra $U(1)^{n-4}$
gauge symmetry, we have to introduce the extra chiral superfields which 
are singlets
under the SM gauge symmetry. Therefore, in the gauge-Higgs unification
scenario, we only unify the gauge fields with two $SU(2)$ Higgs
doublets, not all the Higgs fields. In addition, if we required that
the extra $U(1)^{n-4}$ 
model be chiral, then, we have to introduce the exotic particles
 due to anomaly cancellation.

Another natural question is proton decay. The dangerous tree-level
dimension 5 proton decay operators $[QQQL]_{\theta^2}$ and 
$[UUDE]_{\theta^2}$, and the tree-level Higgs mass term, 
$[H_U H_D]_{\theta^2}$ can be forbidden by $R$ symmetry in general
if one assigned the suitable $R$ charges to the particles~\cite{HN, HNDS}.
At tree-level, the dimension 6 proton decay operators
 by exchange the gauge boson and scalar Higgs 
may be absent, for example, the SM fermions are on the 4-brane
or on the 3-brane which only preserves
$SU(3)\times SU(2)\times U(1)
\times U(1)^{n-4}$ gauge symmetry. However, the proton may decay at
one-loop
level, through the box diagram, for instance, the $SU(5)$ model on
$S^1/(Z_2\times Z_2')$
 with SM fermions in the bulk, the proton can decay through 
box diagram by exchange $X$ and $Y$ in the mean time. So, the GUT
scale can not be much lower than $10^{16}$ GeV~\cite{TJL}.

Because the zero modes preserve $N=1$ supersymmetry, we have to break
the $N=1$ supersymmetry at TeV scale. One approach is
the gaugino mediated supersymmetry breaking, because the SM fermions can
be on the observable 3-brane or 4-brane, the supersymmetry breaking can be
happened 
on the other 3-brane or 4-brane, and can be mediated to the observable
brane by gauge fields and gauginos~\cite{Kaplan}.
 Another approach is $F$-term supersymmetry breaking,
for we will have dilaton and moduli (modulus) in the bulk which
can couple to the SM particles. By the way, in 6-dimension, the complex
$F$-term supersymmetry breaking may induce CP violation, while in
5-dimension, it can not~\cite{HLLY}. 

Furthermore, we can always ask the question: how can we generalize our
procedure? Because if we want to discuss the higher rank GUT symmetry
breaking, for instance, $SU(N)$ $(N > 7)$, $SO(M)$ $(M > 12)$, $E_6$,
$E_7$ and $E_8$,
 we need more $Z_2$ projections. The simple way is that, we introduce
more extra dimensions.

In general, for $N$ supersymmetric GUT with 
rank $n$ gauge group $G$ on the ($4+m$)-dimensional
space-time $M^4\times T^m/(Z_2)^L$, where
$L~\le~ 2m$ for we can have at most
two non-equivalent $Z_2$ reflection symmetries
 along each extra dimension~\cite{TJL}.
 We can choose $L$ $Z_2$ parity operators,
which can be separated into two sets: considering only the zero modes,
one set breaks the supersymmetry and
does not break the gauge symmetry, the other set breaks the gauge group
down 
to $SU(3)\times SU(2) \times U(1)^{n-3}$.
From the set of parity operators which breaks the gauge symmetry,
 we can choose $n-3$ independent
$Z_2$ parity operators $P^{g_1}, P^{g_2}, ..., P^{g_{n-3}}$. Under the
parity
operators $P^{g_1}, P^{g_2}, ..., P^{g_{n-3}}$, the gauge generators
$T^A$ where $A=1, 2, ..., |G|$ for $G$
are separated into $2^{n-3}$ sets: $T^{a_1, a_2, ..., a_{n-3}}$,
$T^{a_1, a_2, ..., {\hat {a}}_{n-3}}$, ...., 
$T^{{\hat {a}}_1, {\hat {a}}_2, ..., {\hat {a}}_{n-3}}$, where $a_i$ and
${\hat {a}}_i$
label the generators in $G/P^{g_i}$ and the coset $G/P^{g_i}$,
respectively, and 
$(a_1, a_2, ..., a_{n-3})$, $(a_1, a_2, ..., {\hat {a}}_{n-3})$, ...,
$({\hat {a}}_1, {\hat {a}}_2, ..., {\hat {a}}_{n-3})$ label the
intersections of those sets.
The explicit examples are given in previous sections.
Under the parities $P^{g_1}, P^{g_2}, ..., P^{g_{n-3}}$, the gauge
generators
transform as 
\begin{eqnarray}
P^{g_i} T^{A_1, ..., a_i, ...A_{n-3}} (P^{g_i})^{-1}
=T^{A_1, ..., a_i, ...A_{n-3}}  ~,~\,
\end{eqnarray}
\begin{equation}
P^{g_i} T^{A_1, ..., {\hat {a}}_i, ...A_{n-3}} (P^{g_i})^{-1}
=-T^{A_1, ..., {\hat{a}}_i, ...A_{n-3}}  ~.~\,
\end{equation}
In addition, one can calculate the KK mode expansions for the superfields,
and then, discuss the particle spectra, the superfields,  the number of
4-dimensional supersymmetries and gauge group on the observable brane or
whole space-time.  

\section{Conclusion}
We study the 6-dimensional 
$N=2$ supersymmetric grand unified theories with gauge
group $SU(N)$ and $ SO(M)$ on the extra space orbifolds $T^2/(Z_2)^3$ and
$T^2/(Z_2)^4$, which can be broken down to 4-dimensional $N=1$ 
$SU(3)\times SU(2)\times U(1)^{n-3}$
gauge symmetry for the zero modes where $n$ is the rank of the gauge
group. We also 
study the models which have two $SU(2)$ Higgs doublets 
(zero modes) from the 6-dimensional vector multiplet. 
For the scenarios without gauge-Higgs unification, we require
that there are no zero modes for
chiral multiplets $\Sigma_5$, $\Sigma_6$ and $\Phi$,
and the zero modes of gauge fields preserve only the
4-dimensional $N=1$ supersymmetry and
$SU(3)\times SU(2)\times U(1)\times U(1)^{n-4}$
gauge symmetry.
For the scenarios with gauge-Higgs unification,
 we require that: (1) there are no zero modes for chiral multiplets
$\Sigma_5$ 
and $\Sigma_6$; (2) considering the zero modes,
there are only one pair of Higgs doublets because if
we had two pairs of Higgs doublets, we may have flavour changing
neutral current problem. With our requirements,
we give the particle spectra, present the fields, the number of
4-dimensional
supersymmetry and gauge group on the 3-brane or
4-brane, and discuss some phenomenology for those models. Furthermore,
we generalize our procedure for $(4+m)$-dimensional $N$ supersymmetric
GUT (with gauge group $G$) breaking on the space-time
 $M^4\times T^m/(Z_2)^L$.

\section*{Acknowledgments}
This research was supported in part by the U.S.~Department of Energy under
 Grant No.~DOE-EY-76-02-3071.

\section*{Appendix A}
For a generic bulk field $\phi(x^{\mu}, y, z)$,
we can define three parity operators $P^y$, $P^z$ and $P^{y' z'}$,
respectively
\begin{eqnarray}
\phi(x^{\mu},y, z)&\to \phi(x^{\mu},-y, z )=P^y \phi(x^{\mu},y, z)
 ~,~\,
\end{eqnarray}
\begin{eqnarray}
\phi(x^{\mu},y, z)&\to \phi(x^{\mu}, y, -z )=P^z \phi(x^{\mu},y, z)
 ~,~\,
\end{eqnarray}
\begin{eqnarray}
\phi(x^{\mu},y', z')&\to \phi(x^{\mu},-y', -z' )= P^{y' z'}
\phi(x^{\mu},y', z')
 ~.~\,
\end{eqnarray}
Denoting the field with ($P^y$, $P^z$, $P^{y' z'}$)=($\pm, \pm, \pm$) by 
$\phi_{\pm \pm \pm}$, we obtain the following KK mode expansions
\begin{eqnarray}
  \phi_{+++} (x^\mu, y, z) &=& 
\sum_{n=0}^{\infty} \sum_{m=0}^{\infty}
 \left( \phi_{+++}^{(2n, 2m)}(x^{\mu}) A_{++}^{2n}(y, R_1) A_{++}^{2m}(z,
R_2)
\right.\nonumber\\&&\left.
\phi^{(2n+1, 2m+1 )}_{+++}(x^\mu) A_{+-}^{2n+1}(y, R_1) A_{+-}^{2m+1}(z,
R_2) \right)
 ~,~\,
\end{eqnarray}
\begin{eqnarray}
  \phi_{++-} (x^\mu, y, z) &=& 
\sum_{n=0}^{\infty} \sum_{m=0}^{\infty}
 \left( \phi_{++-}^{(2n, 2m+1)}(x^{\mu}) A_{++}^{2n}(y, R_1)
A_{+-}^{2m+1}(z, R_2)
\right.\nonumber\\&&\left.
\phi^{(2n+1, 2m )}_{++-}(x^\mu) A_{+-}^{2n+1}(y, R_1) A_{++}^{2m}(z, R_2)
\right)
 ~,~\,
\end{eqnarray}
\begin{eqnarray}
  \phi_{+-+} (x^\mu, y, z) &=& 
\sum_{n=0}^{\infty} \sum_{m=0}^{\infty}
 \left( \phi_{+-+}^{(2n, 2m+1)}(x^{\mu}) A_{++}^{2n}(y, R_1)
A_{-+}^{2m+1}(z, R_2)
\right.\nonumber\\&&\left.
\phi^{(2n+1, 2m+2 )}_{+-+}(x^\mu) A_{+-}^{2n+1}(y, R_1) A_{--}^{2m+2}(z,
R_2) \right)
 ~,~\,
\end{eqnarray}
\begin{eqnarray}
  \phi_{+--} (x^\mu, y, z) &=& 
\sum_{n=0}^{\infty} \sum_{m=0}^{\infty}
 \left( \phi_{+--}^{(2n, 2m+2)}(x^{\mu}) A_{++}^{2n}(y, R_1)
A_{--}^{2m+2}(z, R_2)
\right.\nonumber\\&&\left.
\phi^{(2n+1, 2m+1 )}_{+--}(x^\mu) A_{+-}^{2n+1}(y, R_1) A_{-+}^{2m+1}(z,
R_2) \right)
 ~,~\,
\end{eqnarray}
\begin{eqnarray}
  \phi_{-++} (x^\mu, y, z) &=& 
\sum_{n=0}^{\infty} \sum_{m=0}^{\infty}
 \left( \phi_{-++}^{(2n+1, 2m)}(x^{\mu}) A_{-+}^{2n+1}(y, R_1)
A_{++}^{2m}(z, R_2)
\right.\nonumber\\&&\left.
\phi^{(2n+2, 2m+1 )}_{-++}(x^\mu) A_{--}^{2n+2}(y, R_1) A_{+-}^{2m+1}(z,
R_2) \right)
 ~,~\,
\end{eqnarray}
\begin{eqnarray}
  \phi_{-+-} (x^\mu, y, z) &=& 
\sum_{n=0}^{\infty} \sum_{m=0}^{\infty}
 \left( \phi_{-+-}^{(2n+1, 2m+1)}(x^{\mu}) A_{-+}^{2n+1}(y, R_1)
A_{+-}^{2m+1}(z, R_2)
\right.\nonumber\\&&\left.
\phi^{(2n+2, 2m )}_{-+-}(x^\mu) A_{--}^{2n+2}(y, R_1) A_{++}^{2m}(z, R_2)
\right)
 ~,~\,
\end{eqnarray}
\begin{eqnarray}
  \phi_{--+} (x^\mu, y, z) &=& 
\sum_{n=0}^{\infty} \sum_{m=0}^{\infty}
 \left( \phi_{--+}^{(2n+1, 2m+1)}(x^{\mu}) A_{-+}^{2n+1}(y, R_1)
A_{-+}^{2m+1}(z, R_2)
\right.\nonumber\\&&\left.
\phi^{(2n+2, 2m+2 )}_{--+}(x^\mu) A_{--}^{2n+2}(y, R_1) A_{--}^{2m+2}(z,
R_2) \right)
 ~,~\,
\end{eqnarray}
\begin{eqnarray}
  \phi_{---} (x^\mu, y, z) &=& 
\sum_{n=0}^{\infty} \sum_{m=0}^{\infty}
 \left( \phi_{---}^{(2n+1, 2m+2)}(x^{\mu}) A_{-+}^{2n+1}(y, R_1)
A_{--}^{2m+2}(z, R_2)
\right.\nonumber\\&&\left.
\phi^{(2n+2, 2m+1 )}_{---}(x^\mu) A_{--}^{2n+2}(y, R_1) A_{-+}^{2m+1}(z,
R_2) \right)
 ~,~\,
\end{eqnarray}
where
\begin{eqnarray}
A_{++}^{2n}(y, R_1) = {1 \over\displaystyle {\sqrt{2^{\delta_{n, 0}} \pi
R_1 }}}
 \cos{{2n y}\over {R_1}}
~,~\,
\end{eqnarray}
\begin{eqnarray}
A_{+-}^{2n+1}(y, R_1) = {1 \over\displaystyle {\sqrt{ \pi R_1 }}}
 \cos{{(2n+1)y}\over {R_1}}
~,~\,
\end{eqnarray}
\begin{eqnarray}
A_{-+}^{2n+1}(y, R_1) = {1 \over\displaystyle {\sqrt{ \pi R_1 }}}
 \sin{{(2n+1)y}\over {R_1}}
~,~\,
\end{eqnarray}
\begin{eqnarray}
A_{--}^{2n+2}(y, R_1) = {1 \over\displaystyle {\sqrt{ \pi R_1 }}}
 \sin{{(2n+2)y}\over {R_1}}
~.~\,
\end{eqnarray}
Similarly, we define $A_{++}^{2n}(z, R_2)$, $A_{+-}^{2n+1}(z, R_2)$,
$A_{-+}^{2n+1}(z, R_2)$, $A_{--}^{2n+2}(z, R_2)$.

\section*{Appendix B}
For a generic bulk field $\phi(x^{\mu}, y, z)$,
we can define three parity operators $P^y$, $P^z$ and $P^{y'}$,
respectively
\begin{eqnarray}
\phi(x^{\mu},y, z)&\to \phi(x^{\mu},-y, z )=P^y \phi(x^{\mu},y, z)
 ~,~\,
\end{eqnarray}
\begin{eqnarray}
\phi(x^{\mu},y, z)&\to \phi(x^{\mu}, y, -z )=P^z \phi(x^{\mu},y, z)
 ~,~\,
\end{eqnarray}
\begin{eqnarray}
\phi(x^{\mu},y', z')&\to \phi(x^{\mu},-y', z' )= P^{y'} \phi(x^{\mu}, y',
z')
 ~.~\,
\end{eqnarray}
Denoting the field with ($P^y$, $P^{y'}$, $P^{z}$)=($\pm, \pm, \pm$) by 
$\phi_{\pm \pm \pm}$, we obtain the following KK mode expansions
\begin{eqnarray}
  \phi_{+++} (x^\mu, y, z) &=& 
\sum_{n=0}^{\infty} \sum_{m=0}^{\infty} {1\over {\sqrt{2^{\delta_{m,0}}
\pi R_2}}}
 \phi_{+++}^{(2n, m)}(x^{\mu}) A_{++}^{2n}(y, R_1) \cos{{mz}\over R_2}
 ~,~\,
\end{eqnarray}
\begin{eqnarray}
  \phi_{++-} (x^\mu, y, z) &=& 
\sum_{n=0}^{\infty} \sum_{m=0}^{\infty} {1\over \sqrt{\pi R_2}}
 \phi_{++-}^{(2n, m+1)}(x^{\mu}) A_{++}^{2n}(y, R_1) \sin{{(m+1)z}\over
R_2}
 ~,~\,
\end{eqnarray}
\begin{eqnarray}
  \phi_{+-+} (x^\mu, y, z) &=& 
\sum_{n=0}^{\infty} \sum_{m=0}^{\infty} {1\over {\sqrt{2^{\delta_{m,0}}
\pi R_2}}}
 \phi_{+-+}^{(2n+1, m)}(x^{\mu}) A_{+-}^{2n+1}(y, R_1) \cos{{mz}\over R_2}
~,~\,
\end{eqnarray}
\begin{eqnarray}
  \phi_{+--} (x^\mu, y, z) &=& 
\sum_{n=0}^{\infty} \sum_{m=0}^{\infty} {1\over \sqrt{\pi R_2}}
 \phi_{+--}^{(2n+1, m+1)}(x^{\mu}) A_{+-}^{2n+1}(y, R_1)
\sin{{(m+1)z}\over R_2}
 ~,~\,
\end{eqnarray}

\begin{eqnarray}
  \phi_{-++} (x^\mu, y, z) &=& 
\sum_{n=0}^{\infty} \sum_{m=0}^{\infty} {1\over {\sqrt{2^{\delta_{m,0}}
\pi R_2}}}
 \phi_{-++}^{(2n+1, m)}(x^{\mu}) A_{-+}^{2n+1}(y, R_1) \cos{{mz}\over R_2}
 ~,~\,
\end{eqnarray}
\begin{eqnarray}
  \phi_{-+-} (x^\mu, y, z) &=& 
\sum_{n=0}^{\infty} \sum_{m=0}^{\infty} {1\over \sqrt{\pi R_2}}
 \phi_{-+-}^{(2n+1, m+1)}(x^{\mu}) A_{-+}^{2n+1}(y, R_1)
\sin{{(m+1)z}\over R_2}
 ~,~\,
\end{eqnarray}
\begin{eqnarray}
  \phi_{--+} (x^\mu, y, z) &=& 
\sum_{n=0}^{\infty} \sum_{m=0}^{\infty} {1\over {\sqrt{2^{\delta_{m,0}}
\pi R_2}}}
 \phi_{--+}^{(2n+2, m)}(x^{\mu}) A_{--}^{2n+2}(y, R_1) \cos{{mz}\over R_2}
 ~,~\,
\end{eqnarray}
\begin{eqnarray}
  \phi_{---} (x^\mu, y, z) &=& 
\sum_{n=0}^{\infty} \sum_{m=0}^{\infty} {1\over \sqrt{\pi R_2}}
 \phi_{---}^{(2n+2, m+1)}(x^{\mu}) A_{--}^{2n+2}(y, R_1)
\sin{{(m+1)z}\over R_2}
 ~.~\,
\end{eqnarray}
\section*{Appendix C}
For a generic bulk field $\phi(x^{\mu}, y, z)$,
we can define four parity operators $P^y$, $P^z$, $P^{y'}$ and $P^{z'}$,
respectively
\begin{eqnarray}
\phi(x^{\mu},y, z)&\to \phi(x^{\mu},-y, z )=P^y \phi(x^{\mu},y, z)
 ~,~\,
\end{eqnarray}
\begin{eqnarray}
\phi(x^{\mu},y, z)&\to \phi(x^{\mu}, y, -z )=P^z \phi(x^{\mu},y, z)
 ~,~\,
\end{eqnarray}
\begin{eqnarray}
\phi(x^{\mu},y', z')&\to \phi(x^{\mu},-y', z' )= P^{y'} \phi(x^{\mu},y',
z')
 ~,~\,
\end{eqnarray}
\begin{eqnarray}
\phi(x^{\mu},y', z')&\to \phi(x^{\mu}, y', -z' )= P^{z'} \phi(x^{\mu},y',
z')
 ~.~\,
\end{eqnarray}
Denoting the field with ($P^y$, $P^{y'}$, $P^z$, $P^{ z'}$)=($\pm, \pm,
\pm, \pm$) by 
$\phi_{\pm \pm \pm \pm}$, we obtain the following KK mode expansions
\begin{eqnarray}
\phi_{++++}(x^{\mu}, y, z) = \sum_{n=0}^{\infty} \sum_{m=0}^{\infty}
\phi_{++++}^{(2n, 2m)}(x^{\mu}) A_{++}^{2n} (y, R_1) A_{++}^{2m} (z, R_2) 
 ~,~\,
\end{eqnarray}
\begin{eqnarray}
\phi_{+++-}(x^{\mu}, y, z) = \sum_{n=0}^{\infty} \sum_{m=0}^{\infty}
\phi_{+++-}^{(2n, 2m+1)}(x^{\mu}) A_{++}^{2n} (y, R_1) A_{+-}^{2m+1} (z,
R_2) 
 ~,~\,
\end{eqnarray}
\begin{eqnarray}
\phi_{++-+}(x^{\mu}, y, z) = \sum_{n=0}^{\infty} \sum_{m=0}^{\infty}
\phi_{++-+}^{(2n, 2m+1)}(x^{\mu}) A_{++}^{2n} (y, R_1) A_{-+}^{2m+1} (z,
R_2) 
 ~,~\,
\end{eqnarray}
\begin{eqnarray}
\phi_{++--}(x^{\mu}, y, z) = \sum_{n=0}^{\infty} \sum_{m=0}^{\infty}
\phi_{++--}^{(2n, 2m+2)}(x^{\mu}) A_{++}^{2n} (y, R_1) A_{--}^{2m+2} (z,
R_2) 
 ~,~\,
\end{eqnarray}
\begin{eqnarray}
\phi_{+-++}(x^{\mu}, y, z) = \sum_{n=0}^{\infty} \sum_{m=0}^{\infty}
\phi_{+-++}^{(2n+1, 2m)}(x^{\mu}) A_{+-}^{2n+1} (y, R_1) A_{++}^{2m} (z,
R_2) 
 ~,~\,
\end{eqnarray}
\begin{eqnarray}
\phi_{+-+-}(x^{\mu}, y, z) = \sum_{n=0}^{\infty} \sum_{m=0}^{\infty}
\phi_{+-+-}^{(2n+1, 2m+1)}(x^{\mu}) A_{+-}^{2n+1} (y, R_1) A_{+-}^{2m+1}
(z, R_2) 
 ~,~\,
\end{eqnarray}
\begin{eqnarray}
\phi_{+--+}(x^{\mu}, y, z) = \sum_{n=0}^{\infty} \sum_{m=0}^{\infty}
\phi_{+--+}^{(2n+1, 2m+1)}(x^{\mu}) A_{+-}^{2n+1} (y, R_1) A_{-+}^{2m+1}
(z, R_2) 
 ~,~\,
\end{eqnarray}
\begin{eqnarray}
\phi_{+---}(x^{\mu}, y, z) = \sum_{n=0}^{\infty} \sum_{m=0}^{\infty}
\phi_{+---}^{(2n+1, 2m+2)}(x^{\mu}) A_{+-}^{2n+1} (y, R_1) A_{--}^{2m+2}
(z, R_2) 
 ~,~\,
\end{eqnarray}
\begin{eqnarray}
\phi_{-+++}(x^{\mu}, y, z) = \sum_{n=0}^{\infty} \sum_{m=0}^{\infty}
\phi_{-+++}^{(2n+1, 2m)}(x^{\mu}) A_{-+}^{2n+1} (y, R_1) A_{++}^{2m} (z,
R_2) 
 ~,~\,
\end{eqnarray}
\begin{eqnarray}
\phi_{-++-}(x^{\mu}, y, z) = \sum_{n=0}^{\infty} \sum_{m=0}^{\infty}
\phi_{-++-}^{(2n+1, 2m+1)}(x^{\mu}) A_{-+}^{2n+1} (y, R_1) A_{+-}^{2m+1}
(z, R_2) 
 ~,~\,
\end{eqnarray}
\begin{eqnarray}
\phi_{-+-+}(x^{\mu}, y, z) = \sum_{n=0}^{\infty} \sum_{m=0}^{\infty}
\phi_{-+-+}^{(2n+1, 2m+1)}(x^{\mu}) A_{-+}^{2n+1} (y, R_1) A_{-+}^{2m+1}
(z, R_2) 
 ~,~\,
\end{eqnarray}
\begin{eqnarray}
\phi_{-+--}(x^{\mu}, y, z) = \sum_{n=0}^{\infty} \sum_{m=0}^{\infty}
\phi_{-+--}^{(2n+1, 2m+2)}(x^{\mu}) A_{-+}^{2n+1} (y, R_1) A_{--}^{2m+2}
(z, R_2) 
 ~,~\,
\end{eqnarray}
\begin{eqnarray}
\phi_{--++}(x^{\mu}, y, z) = \sum_{n=0}^{\infty} \sum_{m=0}^{\infty}
\phi_{--++}^{(2n+2, 2m)}(x^{\mu}) A_{--}^{2n+2} (y, R_1) A_{++}^{2m} (z,
R_2) 
 ~,~\,
\end{eqnarray}
\begin{eqnarray}
\phi_{--+-}(x^{\mu}, y, z) = \sum_{n=0}^{\infty} \sum_{m=0}^{\infty}
\phi_{--+-}^{(2n+2, 2m+1)}(x^{\mu}) A_{--}^{2n+2} (y, R_1) A_{+-}^{2m+1}
(z, R_2) 
 ~,~\,
\end{eqnarray}
\begin{eqnarray}
\phi_{---+}(x^{\mu}, y, z) = \sum_{n=0}^{\infty} \sum_{m=0}^{\infty}
\phi_{---+}^{(2n+2, 2m+1)}(x^{\mu}) A_{--}^{2n+2} (y, R_1) A_{-+}^{2m+1}
(z, R_2) 
 ~,~\,
\end{eqnarray}
\begin{eqnarray}
\phi_{----}(x^{\mu}, y, z) = \sum_{n=0}^{\infty} \sum_{m=0}^{\infty}
\phi_{----}^{(2n+2, 2m+2)}(x^{\mu}) A_{--}^{2n+2} (y, R_1) A_{--}^{2m+2}
(z, R_2) 
 ~.~\,
\end{eqnarray}

\end{document}